\RequirePackage{fix-cm}
\documentclass[11pt,a4paper]{article}  

\usepackage{algorithm}
\usepackage{soul}
\usepackage{amssymb}
\usepackage{lineno}
\usepackage{subfig}
\usepackage{array}
\usepackage[export]{adjustbox}
\usepackage{float}
\usepackage{graphicx}
\usepackage{booktabs}
\usepackage{multirow}
\usepackage{amsmath}
\usepackage{xfrac}
\usepackage{commath}
\usepackage{bm}
\usepackage{color,soul}
\usepackage[labelsep=endash]{caption}
\usepackage{makecell}
\usepackage{enumitem}
\usepackage{placeins}
\usepackage{algorithm}
\usepackage{algpseudocode}
\usepackage{rotating}

\newcommand{\uu}{{\bf u}}
\newcommand{\vv}{{\bf v}}
\newcommand{\ww}{{\bf w}}
\newcommand{\zz}{{\bf z}}

\begin{document}

\title{A new fluid-based strategy for the connection of non-matching lattice materials}


\author{Nicola Ferro$^1$, Simona Perotto$^1$, Matteo Gavazzoni$^2$}
\maketitle

\begin{center}
{\small
$^1$
MOX -- 
Dipartimento di Matematica\\
Politecnico di Milano\\
Piazza L. da Vinci, 32, I-20133 Milano, Italy\\
{\tt \{nicola.ferro, simona.perotto\}@polimi.it}
\\[3mm]
$^2$
Dipartimento di Meccanica \\ Politecnico di Milano \\ Via La Masa 1, Milano I-20156, Italy\\
{\tt matteo.gavazzoni@polimi.it}
}
\end{center}
\date{}
\maketitle

\begin{abstract}
We present a new algorithm for the design of the connection region between different lattice materials. We solve a Stokes-type topology optimization problem on a narrow morphing region to smoothly connect two different unit cells.
The proposed procedure turns out to be effective and provides a local re-design of the materials,  leading to a very mild modification of the mechanical behaviour characterizing the original lattices.
The robustness of the algorithm is assessed in terms of sensitivity of the final layout to different parameters. 
Both the cases of Cartesian and non-Cartesian morphing regions are successfully investigated.
\end{abstract}




\section{Introduction}\label{intro}



Cellular materials, commonly known as metamaterials, are artificial structures characterized by the presence of distributed voids in the volume. In particular, structures exhibiting a regular and periodic distribution of voids are referred to as lattice materials. The topology characterizing the Reference Volume Element (RVE), namely the unit cell that is periodically distributed, affects the macroscopic material as a whole. Indeed, different microscopic topologies endow lattices with distinct physical properties. For instance, many lattice materials try to reproduce behaviours which are commonly observed in nature (e.g., wood, sponges, bones), while other microstructural designs aim to mimic uncommon physical characteristics (e.g., auxetic materials). The proposal of newly-conceived lattices is supported also by the spreading of additive manufacturing (AM) techniques, such as $3$D printing~\cite{Thompson2016}.
\\
Lattices are mainly designed via two approaches, either by trial-and-error paradigms to reproduce desired patterns, or by setting suitable optimization problems to guide a rigorous design process. In the second case, inverse homogenization techniques are employed to architect the material distribution in the RVE guaranteeing target properties at the homogenized macroscopic scale~\cite{andreassen2014determine,allaire19,sigmund1994}.
%

In some contexts, it is advisible to carry out a multiscale and/or multimaterial design strategy~\cite{Rodrigues2002,Sanders2021,ArabnejadKhanoki2012}, for instance, in the optimization of components both at a macroscopic (i.e., visible design) and at a microscopic level (i.e., infill characterization), to ensure optimal structural response with respect to some quantities of interest.
In practice, multiscale optimization resorts to lattice materials for the optimal distribution of the microstructured infill and ends up with the identification of several regions inside the macroscopic domain where different materials should be included.
This scenario offers two approaches. On the one hand, the designer can allocate a single-cellular material (i.e., a single topology) whose structural members are appropriately sized in different regions of the macroscopic domain. We associate this framework with functionally graded materials (FGMs)~\cite{Radman2013,Panesar2018}.
On the other hand, it is possible to resort to lattices characterized by different RVE material distribution (i.e., different topologies), thus making the microscopic infill spatially varying in topology and characteristics. The latter approach guarantees more flexibility in the optimal design process as it localizes the use of different materials and, as a consequence, locally diversifies the structural behaviour~\cite{Gao2019,coelho08,Xia2014}.


FGMs and multiple-lattice paradigms massively differ in terms of strategies to deal with heterogeneous unit cells. FGMs exhibit infill patterns which are essentially well-connected due to homogeneity of the involved RVEs. Indeed, the size variation across the domain does not represent a strong issue with a view to manufacturing.
Vice versa, when resorting to different microscopic unit cells, 
two adjacent RVEs are, in general, only partially matching or completely non-matching, thus leading to failing of designs, unfeasible manufacture, and to physics-related issues, such as unwanted stress concentration~\cite{Du2018}. These drawbacks have to be properly tackled and have been addressed in several ways in the recent literature. In~\cite{cramer2016,wang2017}, the authors resort to an approach based on an actual interpolation between two geometries. As an alternative, in~\cite{zhou2008,Radman2013,li2018,zhou2019,zobaer2020,liu2022}, the transition between non matching materials is tackled according to a graded framework, namely, by introducing additional unit cells that implement a progressive morphing of one cell to the other. In general, several cells are involved in such a morphing. Other viable approaches perform a concurrent multiscale topology optimization with explicit matching conditions as a constraint, thus designing at the same time both the macro- and the microscale while enforcing the requested connectivity~\cite{schumacher2015,Du2018,garner2019,liu2020}. This solution is effective, yet computationally heavy as it involves many iterations between the two scales. 

\begin{figure}[h!]
	\centering
	\includegraphics[width=0.6\textwidth]{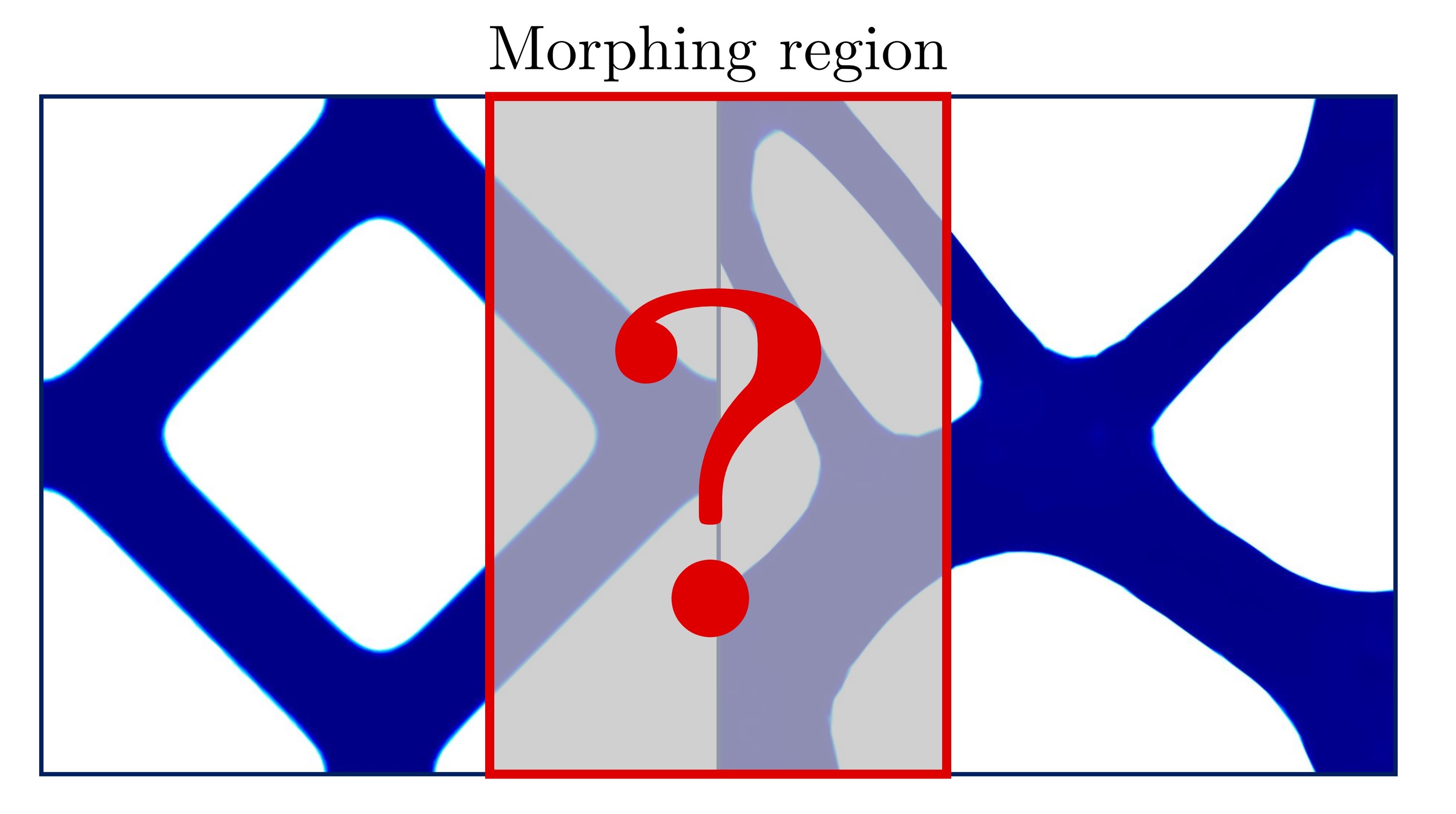}
	\caption{CONFLUENCE algorithm. Sketch of the morphing region when merging different RVEs.}
	\label{figura_intro}
\end{figure}

In this paper, we propose a new methodology, named CONFLUENCE (CONnection by FLUids of differENt CElls), to join different RVEs, which relies on a SIMP-based topology optimization process for fluids~\cite{Borrvall2003}. Starting from two different adjacent unit cells, we identify a morphing region straddling the common side (see Figure~\ref{figura_intro} for a sketch), where we solve a Stokes-based topology optimization, properly constrained by the material distribution in the RVEs to be merged. The design of the matching region is enriched by a customized selection of the computational mesh based on the algorithm SIMP with AnisoTropic mesh adaptivitY (SIMPATY) proposed in~\cite{Micheletti2019,ferro2020density}.

CONFLUENCE algorithm is characterized by two main good features. It has a very localized impact on the cell design since it acts only in a narrow neighbourhood of the side separating the different lattices, in contrast to~\cite{zhou2008,Radman2013,li2018,zhou2019,zobaer2020,liu2022}. Moreover, the mesh adaptation algorithm ensures to 
sharply describe the density across the material/void interface, thus limiting the post-processing phase typical of standard design tools.\\
These features do not strongly affect the original mechanical performance of the joint lattices, ensure a limited computational effort to manage the matching design, and essentially preserve the manufacturability characterizing the considered lattice materials.

The paper is organized as follows. Section~\ref{method} represents the core of the work since providing the new fluid-based methodology to connect diverse lattices, together with the corresponding algorithm. In Section~\ref{results}, we assess CONFLUENCE algorithm on an extensive bunch of test cases, by varying the topology of the considered lattices. Moreover, we investigate the sensitivity of the procedure with respect to the parameters involved in the cell-morphing. Section~\ref{structural_analyses} is devoted to an investigation of the mechanical performance of the joint lattices, with a particular attention to stress distribution and localization. Some conclusions are finally drawn in the last section with a view to future perspectives.


\section{CONFLUENCE algorithm}\label{method}

Connectivity issues among different lattices with non-matching interfaces typically arise when materials characterized by diverse topologies are alternated inside the same component. We consider two adjacent squared and periodic RVEs sharing one edge. The goal is to set a design procedure to join different lattices in a continuous and smooth fashion, by modifying the original topologies in a narrow neighbourhood of the common side, that we name morphing region. To this end, we solve in such an area a density-based topology optimization problem. In particular, we consider a fluid-type problem, completed with appropriate boundary conditions on the velocity profile and on the density in order to be compliant with the original topologies of the matched lattices. Throughout the paper, standard notations for the function spaces are employed~\cite{ern04}.

\subsection{Stokes flow-driven topology optimization}
We consider a flow design problem which describes the distribution of porous and impermeable material in a domain $Y$, here coinciding with the morphing region. In more detail, the design phase is driven by a Stokes-type equation, subject to given constraints and targeting a goal functional~\cite{Borrvall2003}. Following a density-based approach, we resort to the density (or design) variable $\rho \in L^\infty(Y, [0, 1])$, which identifies the topology under optimization. We associate $\rho = 0$ with the impermeable material (i.e., the solid), whereas $\rho = 1$ characterizes the fully porous regions (i.e., the fluid).

To clearly formalize the constrained optimization problem, we first introduce the weak form of the (generalized) Stokes equation for the velocity, $\uu$, and the pressure, $p$, i.e.,
\\
Find $(\uu, p) \in U_{\bf g} \times Q$ such that
\begin{equation}\label{stokes_initial}
\left\{
\begin{array}{lll}
& a(\uu,\vv) +  b(\vv, p) = F(\vv) \quad & \forall \vv \in U,
\\[2mm]
& b(\uu, q) = 0 \quad & \forall q \in Q,
\end{array}
\right.
\end{equation}
where
\begin{equation}\label{stokesa}
a(\ww, \zz) = \int_{Y} \mu \nabla \ww : \nabla \zz dY + \int_{Y} \alpha \ww \cdot \zz dY,
\end{equation}
$$
b(\zz, r) = \int_{Y} -r \nabla \cdot \zz dY, \quad F(\zz) = \int_{Y} {\bf f} \cdot \zz dY,
$$
are the Stokes bilinear ($a(\cdot, \cdot)$ and $b(\cdot, \cdot)$) and linear ($F(\cdot)$) forms,  for any $\ww \in U_{\bf g}$, $\zz \in U$ and for any $r \in Q$.
Here, we are assuming to close problem~\eqref{stokes_initial} with a Dirichlet data, $\uu = {\bf g}$, for the velocity on the boundary portion $\Gamma_D \subset \partial Y$, so that function spaces $U_{\bf g}$, $U$ and $Q$ can be selected as 
$U_{\bf g} = \{ \vv \in [H^1(Y)]^2, \vv \lvert_{\Gamma_D} = {\bf g} \}$,
$U = \{ \vv \in [H^1(Y)]^2, \vv \lvert_{\Gamma_D} = {\bf 0} \}$, $Q = L^2(Y)$.
Moreover, ${\bf f} \in [L^2(Y)]^2$ denotes an external forcing term, $\mu \in \mathbb{R}_+$ the diffusivity of the fluid, and $\alpha \in \mathbb{R}_+$ the inverse permeability of the considered medium.

In~\cite{Borrvall2003}, the bilinear form~\eqref{stokesa} is modified to account for the presence of the design variable. Thus, we replace the form $a(\cdot, \cdot)$ with
$$
a_\rho(\ww, \zz) = \int_{Y} \mu \nabla \ww : \nabla \zz dY + \int_{Y} \alpha_\rho \ww \cdot \zz dY,
$$
where the constant inverse permeability $\alpha$ is now substituted by the function of $\rho$
\begin{equation}\label{alpha_range}
\alpha_\rho = \alpha_\rho(\rho) = \overline{\alpha} + (\underline{\alpha} - \overline{\alpha}) \, \rho \, \dfrac{1 + \phi}{\rho + \phi},
\end{equation}
with $\overline{\alpha}$ and $\underline{\alpha} \in \mathbb{R}_+$ the upper and lower bound for the inverse permeability, respectively. The scalar $\phi > 0$ is a penalization parameter, which strongly promotes a sharp alternation of porous and impermeable materials for large values. We remark that $\alpha_\rho$ is equal to $\overline{\alpha}$ for $\rho = 0$ and to $\underline{\alpha}$ for $\rho = 1$. Thus, the regions characterized by $\rho = 0$ have high inverse permeability (i.e., low permeability), and correspond to solid material; vice versa regions where $\rho=1$ are associated with the fluid.


The topology optimization problem for the allocation of solid and fluid regions reads
\begin{equation}\label{min_stokes}
\min_{\rho\in L^\infty(Y)} \mathcal{J}(\uu, p, \rho) :
\left\{
\begin{array}{l}
\begin{array}{ll}
a_\rho(\uu,\vv) +  b(\vv, p) = F(\vv) \quad 
& \forall \vv \in U 
\\[2mm]
b(\uu, q) = 0 & \forall q \in Q
\end{array}
\\[7mm]
\displaystyle\int_Y \rho dY \le \beta \lvert Y \rvert
\\[3mm]
0 \le \rho \le 1,
\end{array}
\right.
\end{equation}
where $\mathcal{J}(\uu, p, \rho)$ is the selected objective functional to be minimized; the two equations enforce the Stokes regime, with $\uu \in U_{\bf g}$ and $p \in Q$; the first inequality prescribes the maximum fraction $\beta \lvert Y \lvert$ of fluid phase to be allocated in the domain $Y$, with
$\beta \in (0, 1)$ and $\lvert Y \lvert$ the domain measure; the last box constraint keeps trace of the range prescribed to the design variable $\rho$.
In the sequel, we adopt the total potential energy of the fluid
\begin{equation}\label{goal}
\mathcal{J}(\uu, p, \rho) = \dfrac{1}{2} a_\rho(\uu, \uu) - F(\uu),
\end{equation}
as objective functional. We observe that problem \eqref{min_stokes} does not suffer from uniqueness issues~\cite{Borrvall2003} (as well as from drawbacks related to the numerical approximation), 
differently from the case when the optimization process is constrained by the linear elasticity equation~\cite{Bendsoe2004}.

\subsection{A fluid-based approach to connect non-matching lattices}

\begin{figure}[h!]
	\centering
	\includegraphics[width=0.6\textwidth]{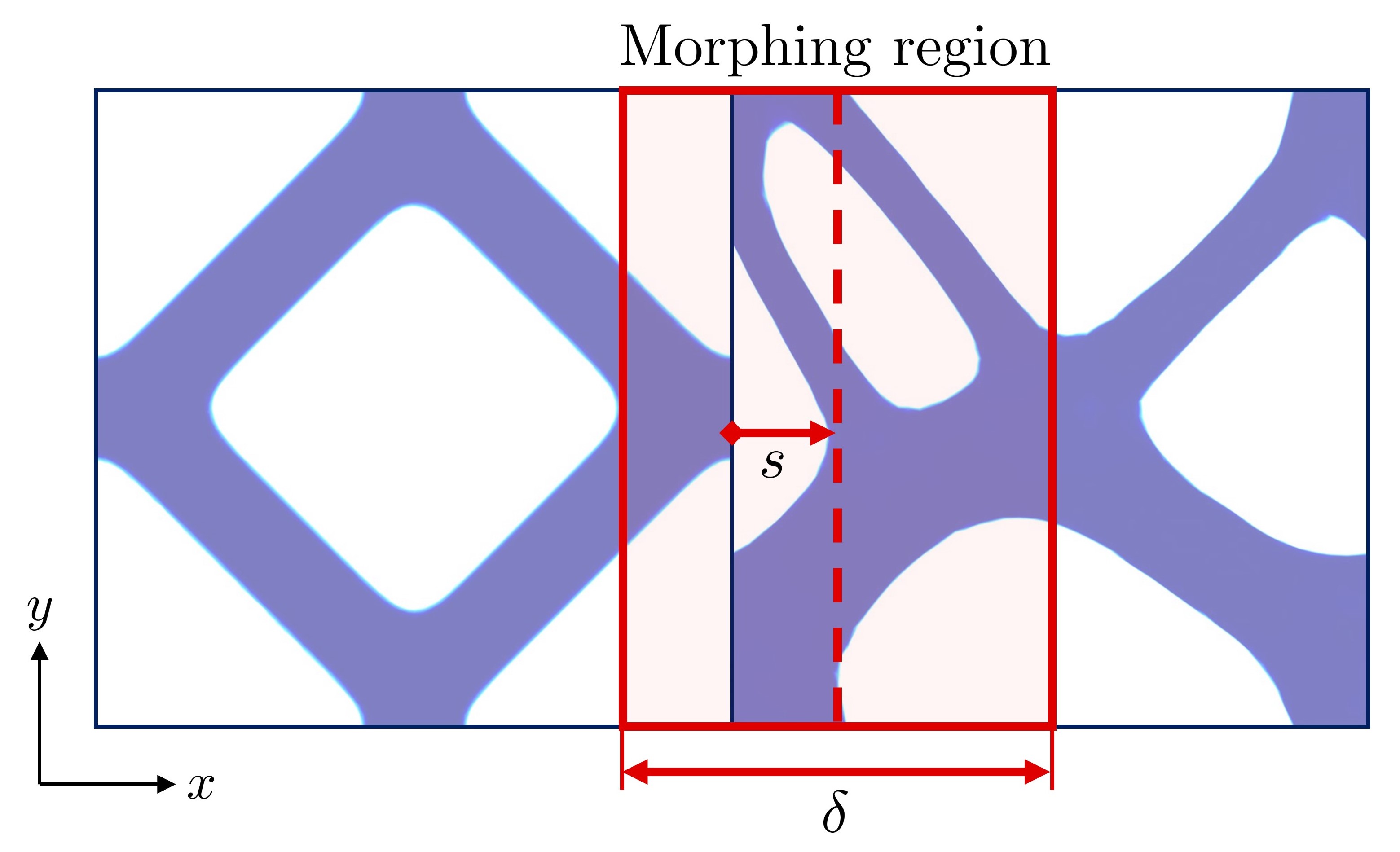}
	\caption{Morphing region. Definition of the main geometric parameters.}
	\label{morphing_regione}
\end{figure}

We exploit the topology optimization problem presented in the previous section to design the transition area from one lattice material to an adjacent one. 
As a reference setting, we consider two design domains, $Y_L$ and $Y_R$, which share the common (entire) side $\mathcal E = \{(x_{\mathcal{E}}, y): y_l \le y \le y_u\}$, with $x_{\mathcal{E}}$, $y_l$, $y_u \in \mathbb{R}$. The optimization process takes place in the rectangular morphing region $Y = ( x_{\mathcal{E}} + s  - \delta/2, x_{\mathcal{E}} + s  + \delta/2 ) \times (y_l, y_u)$, with $s \in \mathbb{R}$ the shift and $\delta \in \mathbb{R}_+$ the width of the morphing region (see Figure \ref{morphing_regione}).

In order to merge the two lattices associated with $Y_L$ and $Y_R$, problem \eqref{min_stokes} is solved in the morphing region, after prescribing ad hoc boundary conditions on $\uu$ and $\rho$.
We label the two vertical sides of the morphing region boundary $\partial Y$ by $\Gamma_L = \partial Y \cap Y_L$ and $\Gamma_R = \partial Y \cap Y_R$, which are instrumental to set the minimization in problem \eqref{min_stokes}.
\\
The proposed procedure can be itemized as follows:
\begin{enumerate}
    \item[i)] we read as an input the densities $\rho_L \in L^\infty(Y_L, [0, 1])$ and $\rho_R \in L^\infty(Y_R, [0, 1])$ identifying the original unit cell topologies to be merged;
    \item[ii)] we solve the generalized Stokes problems
\\
Find $(\uu_i, p_i) \in U_i \times Q_i$ such that
\begin{equation}\label{auxiliary}
\left\{
\begin{array}{ll}
a_{\rho_i}(\uu_i,\vv) + b(\vv, p_i) = F_i(\vv) \quad & \forall \vv \in U_i
\\[2mm]
b(\uu_i, q) = 0 \quad & \forall q \in Q_i,
\end{array}
\right.
\end{equation}
where $i = L, R$, $U_i = \{ \vv \in [H^1(Y_i)]^2\}$, $Q_i = \{q \in L^2(Y_i)\}$, and $F_i(\vv) = \int_{Y_i} {\bf f}_i \cdot \vv dY_i$, with ${\bf f}_i$ a forcing term orthogonal to the side $\mathcal E$;
\item[iii)] we define the boundary conditions to the optimization problem \eqref{min_stokes} by defining the two Dirichlet data ${\bf g}_{in} = \uu_L \lvert_{\Gamma_L}$ and ${\bf g}_{out} = \uu_R \lvert_{\Gamma_R}$;
\item[iv)] we solve in the morphing region the topology optimization problem
\begin{equation}\label{min_topopt}
\hspace*{-0.5cm}
\min_{\rho\in L^\infty(Y)} \mathcal{J}(\uu, p, \rho) :
\left\{
\begin{array}{l}
\begin{array}{ll}
a_\rho(\uu,\vv) + b(\vv, p)  = F(\vv)
\quad & \forall \vv \in W 
\\[2mm]
b(\uu, q) = 0 & \forall q \in Q
\end{array}
\\[3mm]
\rho \lvert_{\Gamma_L} = \rho_L \lvert_{\Gamma_L}
\\[3mm]
\rho \lvert_{\Gamma_R} = \rho_R \lvert_{\Gamma_R}
\\[3mm]
\displaystyle\int_Y \rho dY \le \beta \lvert Y \rvert
\\[3mm]
0 \le \rho \le 1,
\end{array}
\right.
\end{equation}
where 
$W = \{ \ww \in [H^1_\#(Y)]^2 : \ww \lvert_{\Gamma_L \cup \Gamma_R} = {\bf 0} \},
$ with $\uu \in W_{\bf g} = \{ \ww \in [H^1_\#(Y)]^2 : \ww \lvert_{\Gamma_L} = {\bf g}_{in},  \ww \lvert_{\Gamma_R} = {\bf g}_{out} \}$,
and where $H^1_\#(Y)$ is the space of the $H^1(Y)$-functions which are periodic along $\partial Y \setminus (\Gamma_L \cup \Gamma_R)$.

\end{enumerate}


We remark that the data assigned to the density on $\Gamma_L$ and $\Gamma_R$ and the essential boundary conditions characterizing the space $W_{\bf g}$ have a different, albeit complementary, role in the design of the transition topology. In particular, the former enforce the density continuity along the vertical sides of the morphing region, acting as a gluing expedient; the latter impose a smooth morphing between the original and the transition topologies, so that no sharp features characterize the junction. As confirmed by the numerical assessment, these boundary assignments have to be simultaneously imposed in order to guarantee a seamless transition design.

\subsection{Numerical discretization}\label{numerical_discretization}
The numerical implementation of problem~\eqref{min_topopt} is tackled in a continuous finite element setting~\cite{ern04}.
For this purpose, we introduce the computational mesh $\mathcal{T}_h(\Omega) = \{K\}$, a conforming triangular tessellation associated with the generic domain $\Omega$, and the corresponding discrete space of continuous piecewise polynomials of degree $s \in \mathbb{N}_+$,
$$
X_h^s(\Omega) = \big\{ v \in C^0(\overline{\Omega}) : v \lvert_K \in \mathbb{P}_s(K) \ \forall K \in \mathcal{T}_h(\Omega) \big\}.
$$
Within this framework, we approximate problems \eqref{auxiliary} to discretize $\uu_{L}$ and $\uu_{R}$ instrumental to assign the boundary data in the discrete counterpart of problem \eqref{min_topopt}. Thus, we solve
\\
Find $(\uu_{h,i}, p_{h,i}) \in U_{h,i} \times Q_{h,i}$ such that
\begin{equation}\label{auxiliary_discrete}
\left\{
\hspace*{-0.25cm}
\begin{array}{lll}
& a_{\rho_{h,i}}(\uu_{h,i},\vv_h) +  b(\vv_h, p_{h,i}) = F_i(\vv_h) \, & \forall \vv_h \in U_{h,i},
\\[2mm]
& b(\uu_{h,i}, q_h) = 0 \, & \forall q_h \in Q_{h,i},
\end{array}
\right.
\end{equation}
with $U_{h,i} = [X_h^2(Y_i)]^2$ and $Q_{h,i} = X_h^1(Y_i)$ and with $\rho_{h,i}$ the discrete density in $X_h^1(Y_i)$, for $i = L, R$.
The spaces $U_{h,i}$ and $Q_{h,i}$ ensure the inf-sup condition, i.e., the well-posedness of problems~\eqref{auxiliary_discrete} and the absence of spurious oscillations in the discrete solutions $(\uu_{h,i}, p_{h,i})$~\cite{brezzi2013}.

Successively, problem \eqref{min_topopt} is tackled by resorting to a gradient-based optimizer for the minimization of the functional and to a finite element scheme to approximate the state equations.
The discrete counterpart of the topology optimization problem \eqref{min_topopt} reads 

\begin{equation}\label{min_topopt_discrete}
\min_{\rho_h\in V_h} \mathcal{J}(\uu_h, p_h, \rho_h):
\left\{
\begin{array}{l}
\begin{array}{ll}
a_{\rho_h}(\uu_h,\vv_h) + b(\vv_h, p_h) 
 = F(\vv_h) \quad & \forall \vv_h \in W_h 
\\[3mm]
b(\uu_h, q_h) = 0 & \forall q_h \in Q_h
\end{array}
\\[3mm]
\rho_h \lvert_{\Gamma_L} = \rho_{h,L} \lvert_{\Gamma_L}
\\[3mm]
\rho_h \lvert_{\Gamma_R} = \rho_{h,R} \lvert_{\Gamma_R}
\\[3mm]
\displaystyle\int_Y \rho_h dY \le \beta \lvert Y \rvert
\\[3mm]
0 \le \rho_h \le 1,
\end{array}
\right.
\end{equation}
with $V_h = X_{\#, h}^1(Y)$, $Q_h = X_h^1(Y)$,
$W_h = \{\ww_h \in [X_{\#, h}^2(Y)]^2, \, \ww \lvert_{\Gamma_i} = {\bf 0}, \, i = L, R \}$,  where
$\uu_h \in W_{{\bf g}, h} = \{\ww_h \in [X_{\#, h}^2(Y)]^2, \, \ww_h \lvert_{\Gamma_L} = {\bf g}_{in, h}, \ww_h \lvert_{\Gamma_R} = {\bf g}_{out, h}\}$, with ${\bf g}_{in, h}$ and ${\bf g}_{out, h}$ suitable finite element approximations in $[X_h^2(Y)]^2$ of the data ${\bf g}_{in}$ and ${\bf g}_{out}$, and where symbol $\#$ refers to the periodicity of the functions along $\partial Y \setminus (\Gamma_L \cup \Gamma_R)$.

We solve problem \eqref{min_topopt_discrete} by means of a variant of the SIMPATY (SIMP with Anisotropic adaptiviTY) algorithm introduced in~\cite{Micheletti2019} and successfully validated in different application settings~\cite{ferro2020density,Ferro2020a,Ferro2020b,Ferro2021,Cristofaro2021,ferro19}.
SIMPATY algorithm efficiently combines the well-established SIMP method for topology optimization~\cite{Bendsoe2004} with an advanced mesh adaptation technique. 
An a posteriori estimator for the discretization error associated with the density variable is exploited to drive a metric-based mesh adaptation process~\cite{Frey2008}. In particular, the authors resort to an anisotropic counterpart (formulated in~\cite{enumath09}) of the well-known recovery-based error analysis proposed by O.C. Zienkiewicz and J.Z. Zhu in~\cite{ZZ1987,Zienkiewicz1992}. 
The employment of anisotropic meshes allow to strike a balance between accuracy and efficiency as it is consolidated in the literature~\cite{Dompierre2002,perotto2015,perotto2022}. 

As a matter of fact, SIMPATY algorithm ensures to design high-quality optimized layouts, characterized by smooth boundaries and free-form features, while healing some drawbacks typical of density-based topology optimization approaches~\cite{filtrati,SP98}, such as the staircase effect and the presence of too thin details, unpractical for manufacturing. Successively, in order to make mechanical analysis free from any bias induced by stretched elements, 
a hybrid version of SIMPATY algorithm has been proposed in~\cite{Ferro2020b}. Here, the idea is to
prescribe sufficiently small isotropic elements in correspondence with the internal portion of the structures ($\rho \simeq 1$), while preserving deformed triangles in order to sharply detect the layout boundary.

Although topology optimization based on a Stokes flow exhibits, in general, few drawbacks related to the selected grid, we solve problem \eqref{min_topopt_discrete} on a sequence of anisotropic adapted meshes to benefit of the computational advantages led by such a technique.\\
The whole procedure is listed in the algorithm below.

\begin{algorithm}[H]
	\caption{: CONFLUENCE (CONnection by FLUids of differENt CElls)}\label{algo}
	{\bf Input}:
	${\tt CTOL, TOL, TOPT, kmax}$, $\rho_{h}^{\tt 0}$,  $\mathcal{T}_h^{{\tt 0}}$, $\beta$, $\text{RVE}_L$, $\text{RVE}_R$, $Y$
	\begin{algorithmic}[1]
		\State Set: ${\tt k}$ = 0, errC $= 1+{\tt CTOL}$;
		\vspace{1mm}
		\State $[\rho_{h,L}; Y_L] = {\tt import}(\text{RVE}_L)$;\label{algo:dataL}
		\vspace{1mm}
		\State $[\rho_{h,R}; Y_R] = {\tt import}(\text{RVE}_R)$;\label{algo:dataR}
		\vspace{1mm}
		\State $\uu_{h,L} = {\tt solve\_flux}(Y_L, \rho_{h,L})$;\label{algo:fluxL}
		\vspace{1mm}
		\State $\uu_{h,R} = {\tt solve\_flux}(Y_R, \rho_{h,R})$;\label{algo:fluxR}
		\vspace{1mm}
		\State ${\tt mc}= {\tt assign}(\Gamma_L, \uu_{h,L}, \rho_{h,L}; \Gamma_R, \uu_{h,R}, \rho_{h,R})$;\label{algo:mc}
		\While {errC $> {\tt CTOL}$ \&  ${\tt k} < {\tt kmax}$}\label{algo:while}
		\vspace{1mm}
		\State $\rho_{h}^{{\tt k+1}}$ = ${\tt optimize}(\mathcal{J}, \nabla_{\rho}           \mathcal{J}, \beta, {\tt mc}, \rho_{h}^{\tt k}, {\tt TOPT})$;\label{algo:opti}
		\State $\mathcal{T}_h^{{\tt k+1}}$ = ${\tt adapt}(\mathcal{T}_h^{{\tt k}}$ , $\rho_{h}^{{\tt k+1}}, {\tt TOL}$);\label{algo:adapt}
		\vspace{1mm}
		\State errC = $\lvert \#\mathcal{T}_h^{{\tt k}+1}- \#\mathcal{T}_h^{{\tt k}}\rvert/\#\mathcal{T}_h^{{\tt k}}$;
		\vspace{1mm}
		\State ${\tt k} = {\tt k}+1$;
		\EndWhile \label{algo:endwhile}
		\vspace{1mm}
		\State $\rho_Y = \rho_{h}^{{\tt k}}$;
		\State [$\rho_{LYR}$; $\Omega^\circ$]	= {\tt join}($\rho_{h, L}$, $Y_L$, $\rho_{h, R}$, $Y_R$, $\rho_Y$, Y);\label{algo:join}
			\end{algorithmic}
	\vspace{1mm}
	{\bf Output}: $\rho_Y$, $\Omega^\circ$,  $\rho_{LYR}$
\end{algorithm}
The first phase which occurs is a pre-processing step (lines \ref{algo:dataL}-\ref{algo:mc}). In particular, the left and right RVEs are read by function $\tt import$ in terms of density function and design domain (lines \ref{algo:dataL}-\ref{algo:dataR}). Successively, the auxiliary problems in \eqref{auxiliary_discrete} are solved to compute the approximations $\uu_{h,L}$ and $\uu_{h,R}$ (lines \ref{algo:fluxL}-\ref{algo:fluxR}), which, together with the associated densities, $\rho_{h,L}$ and $\rho_{h,R}$, are employed to assemble the matching conditions, ${\tt mc}$, on $\Gamma_L$ and $\Gamma_R$ through the function ${\tt assign}$ (line \ref{algo:mc}).
\\
The main optimization loop consists of lines \ref{algo:while}-\ref{algo:endwhile}. At each iteration $\tt k$, ${\tt optimize}$ performs the topology optimization in \eqref{min_topopt_discrete} in a $\xi_{\tt k}$ number of iterations. Such routine is fed with the goal functional, $\mathcal J$, the associated gradient with respect to the density, $\nabla_\rho \mathcal{J}$, the volume fraction $\beta$, the matching conditions $\tt mc$ on $\Gamma_L$ and $\Gamma_R$, the distribution of the density at the previous iteration, $\rho_{h}^{\tt k}$, and the tolerance ${\tt TOPT}$ to break the optimization (line \ref{algo:opti}). In particular, with reference to the gradient of the functional $\nabla_\rho \mathcal{J}$, we employ a standard Lagrangian approach~\cite{Lions1971}. As an alternative, one may resort to automatic differentiation~\cite{margossian2019}.
\\
Finally, the optimized density is employed to drive an anisotropic mesh adaptation (line \ref{algo:adapt}). In particular, function $\tt adapt$ minimizes the number of the mesh elements for a fixed accuracy $\tt TOL$ on the discretization error for the density. To this aim, the size, shape, and orientation of the triangles are properly tuned, jointly with an error equidistribution criterion (we refer the interested reader to \cite{Bolla06,Micheletti2010} for all the details).
Two criteria constrain the {\bf while} loop, the first one in order to control the stagnation of the mesh cardinality, the second one to ensure a termination within a $\tt kmax$ number of iterations. 

The algorithm eventually delivers the final layout $\rho_Y$ that describes the optimized topology in the morphing region $Y$; the whole design domain $\Omega^\circ$, with $\Omega = \overline{Y}_L \cup \overline{Y}_R$ and where $^\circ$ denotes the internal part of the associated set; the distribution of the density on $\Omega^\circ$, given by
$$\rho_{LYR}({\bf x} ) = 
\left\{
\begin{array}{lll}
\rho_{h, L} ({\bf x} )
& \quad & {\bf x} \in Y_L \setminus Y
\\[2mm]
\rho ({\bf x} )
& & {\bf x} \in \overline{Y}
\\[2mm]
\rho_{h, R} ({\bf x} )
& & {\bf x} \in Y_R \setminus Y.
\end{array}
\right.
$$
The global density, $\rho_{LYR}$, and domain, $\Omega^\circ$, are assembled by function $\tt join$ in line~\ref{algo:join}.

\section{Results}\label{results}
\begin{figure}[h!]
	\centering
	\includegraphics[width=0.65\textwidth]{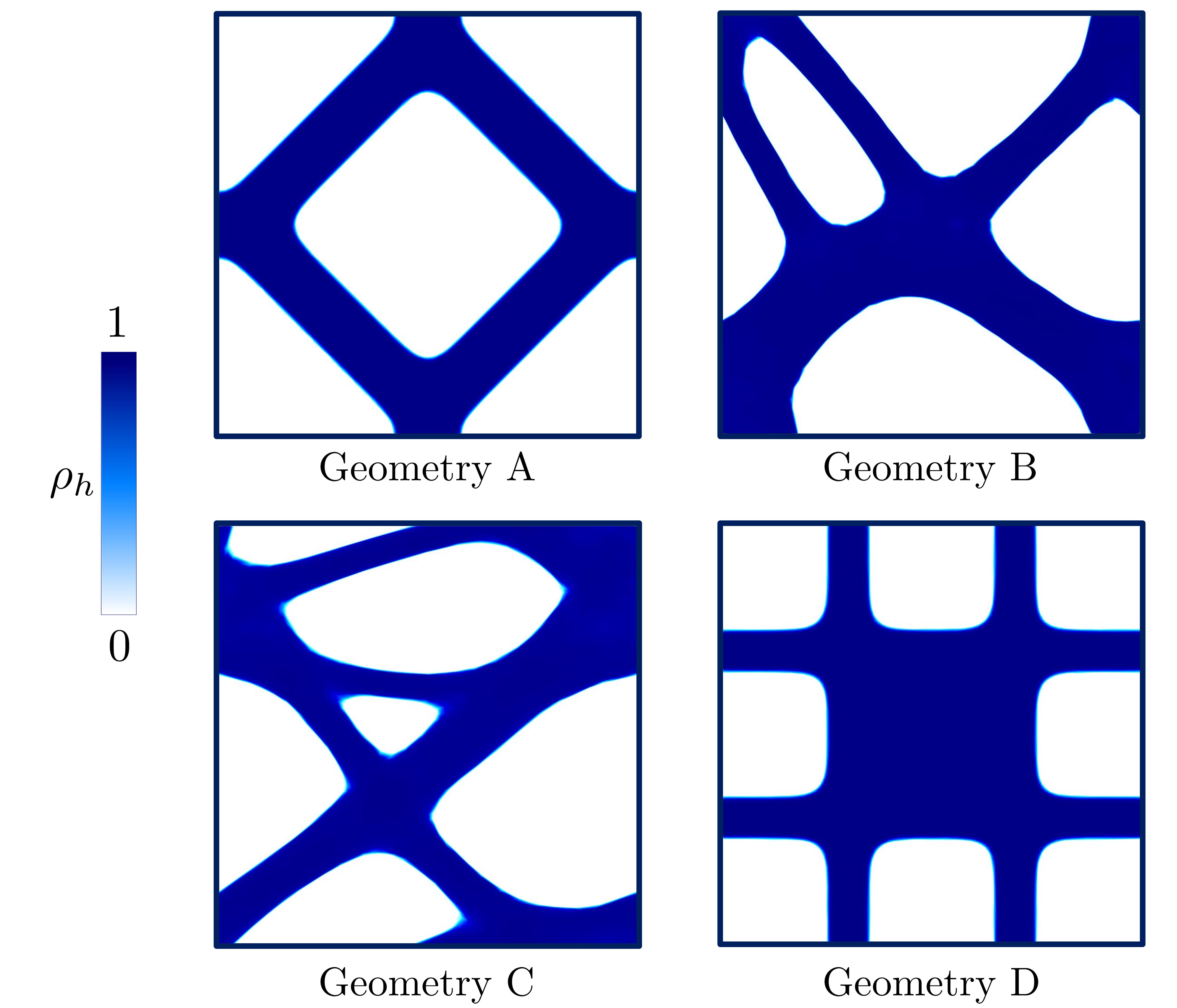}
	\caption{Geometries used to assess the effectiveness of CONFLUENCE algorithm.}
	\label{geometries}
\end{figure}

The effectiveness of CONFLUENCE algorithm for connecting non-matching lattice materials is here investigated, with emphasis on the role played by the main parameters involved in the design strategy in \eqref{min_topopt_discrete}. 
As a benchmark scenario, we join two RVEs, starting from the simplified configuration in Figure~\ref{morphing_regione} where the common edge $\mathcal E$ is vertical and the morphing region does coincide with a Cartesian domain. As different  geometries to be connected, we consider the four  topologies in Figure~\ref{geometries} associated with the square $(0,1)^2$, namely two standard unit cells available in the literature (geometries A and D), and two free-form layouts (geometries B and C) yielded by the extension of the SIMPATY algorithm to the design of lattice materials~\cite{ferro2020density}. Such design tool, named microSIMPATY algorithm, relies on SIMPATY procedure and on a standard inverse homogenization approach, to propose new unit cells which confer prescribed (homogenized) physical properties at the macroscale. For instance, geometries B and C have been designed in a thermo-elastic multiobjective context, to ensure a high shear stiffness while exhibiting an isotropic (geometry B) or an anisotropic (geometry C) thermal and stiffness behaviour~\cite{Gavazzoni2022}.
\\
All the unit cells in Figure~\ref{geometries} are characterized by a poor one-to-one connectivity, so that appropriate junctions are advisable.

For all the test configurations in this section, we preserve the same choices for some of the quantities tuning CONFLUENCE morphing. We start the optimization problem on an initial unstructured grid, $\mathcal{T}_h^0$, characterized by a uniform spacing $h \simeq 1/60$, and selecting an initial constant $\rho_h^{\tt 0} = 0.5$; we set the penalization parameter $\phi$ to $0.6$, the diffusivity $\mu$ to $1$, the lower bound for the inverse permeability $\underline{\alpha}$ to $2.5 \, \mu \cdot 10^{-4}$, ${\bf f}$ to $[1, 0]^T$,
${\tt CTOL}$ to $1.5 \cdot 10^{-2}$, ${\tt TOL}$ to $2.5\cdot 10^{-6}$, ${\tt TOPT}$ to $10^{-3}$, $\tt kmax$ to $50$.
Moreover, to perform the connection between two heterogeneous cells, we identify $Y_L$ with the unit square $(-1, 0) \times (0, 1)$ and $Y_R$ with the unit square $(0, 1)^2$.
Then, the volume fraction $\beta$ in \eqref{min_topopt} is chosen as
$$
\beta = \displaystyle \frac{1}{\lvert Y \rvert} \left (\int_{Y_L \cap Y} \rho_{h, L} dY + \int_{Y_R \cap Y} \rho_{h, R} dY \right),
$$
in order to preserve the quantity of material in $Y$ during the morphing process.\\
Finally, concerning function $\tt optimize$ in Algorithm~\ref{algo}, we select the gradient-based optimizer package IPOPT~\cite{Waechter2006}, while function $\tt adapt$ resorts to the metric-based mesh generator BAMG (Bidimensional Anisotropic Mesh Generator). Both these computational tools are
embedded in $\tt FreeFEM$~\cite{FreeFem} which is the solver we adopt.

\paragraph{Matching geometries A and B}

\begin{figure}[h]
	\centering
	\includegraphics[width=0.55\textwidth]{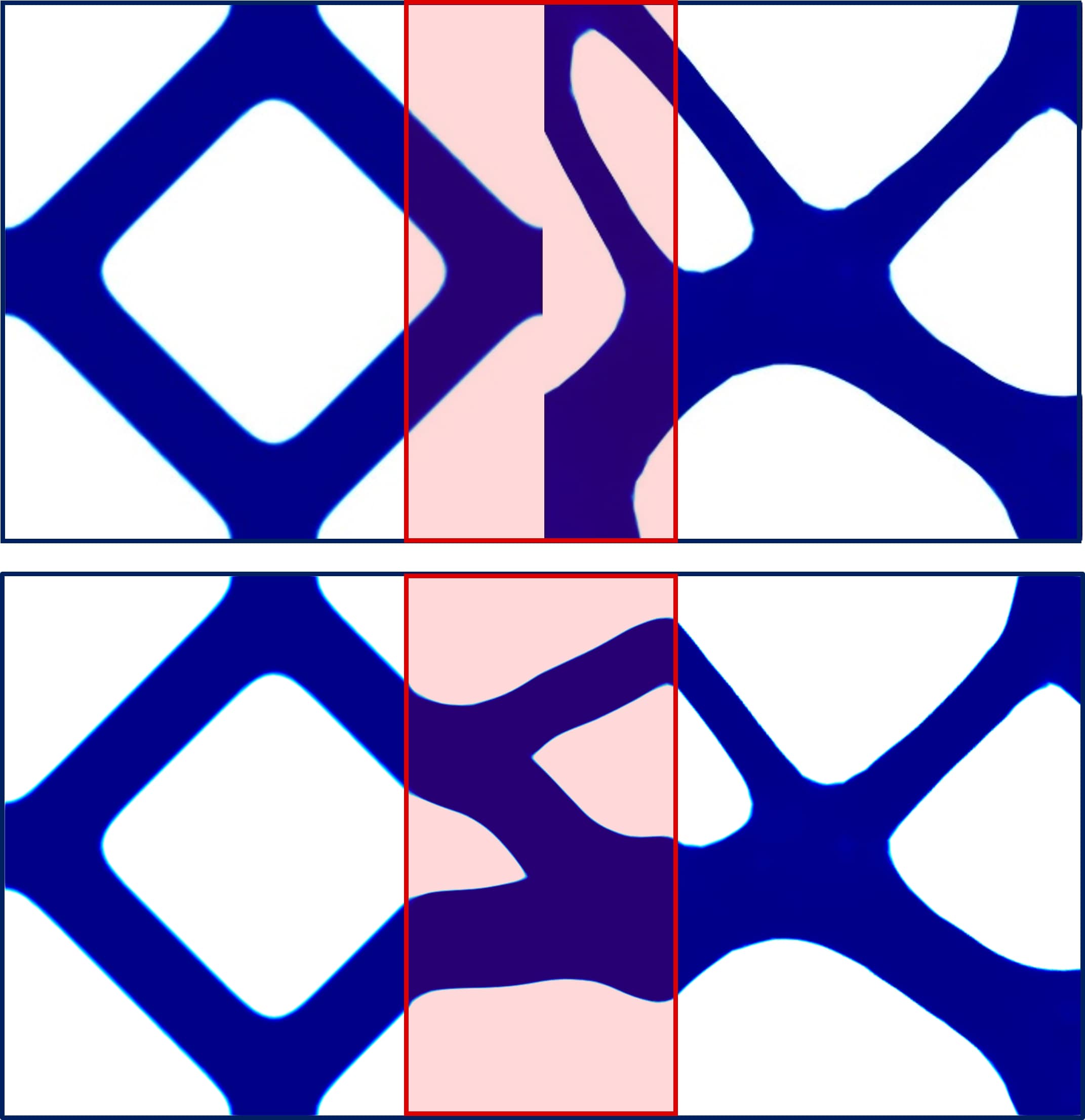}
	\caption{Matching A-B.  Initial (top) and  final (bottom) density distribution provided by CONFLUENCE in $\Omega^\circ$.}
	\label{square_sim1}
\end{figure}

The first scenario used to assess CONFLUENCE morphing involves geometries A (on the left) and B (on the right). As shown in Figure~\ref{square_sim1} (top), the cells to be joined have a null intersection along the common interface $\mathcal E$. This feature challenges the proposed procedure to identify the new topology in the morphing region. We select $Y$ so that $s = 0$ (i.e., the region is centered at $\mathcal E$) and $\delta = 0.5$ (see Figure~\ref{morphing_regione}). Finally, for this configuration, we set $\overline{\alpha} = 2.5 \, \mu \cdot 10^5$. 

\begin{figure}[H]
	\centering
	\includegraphics[width=0.7\textwidth]{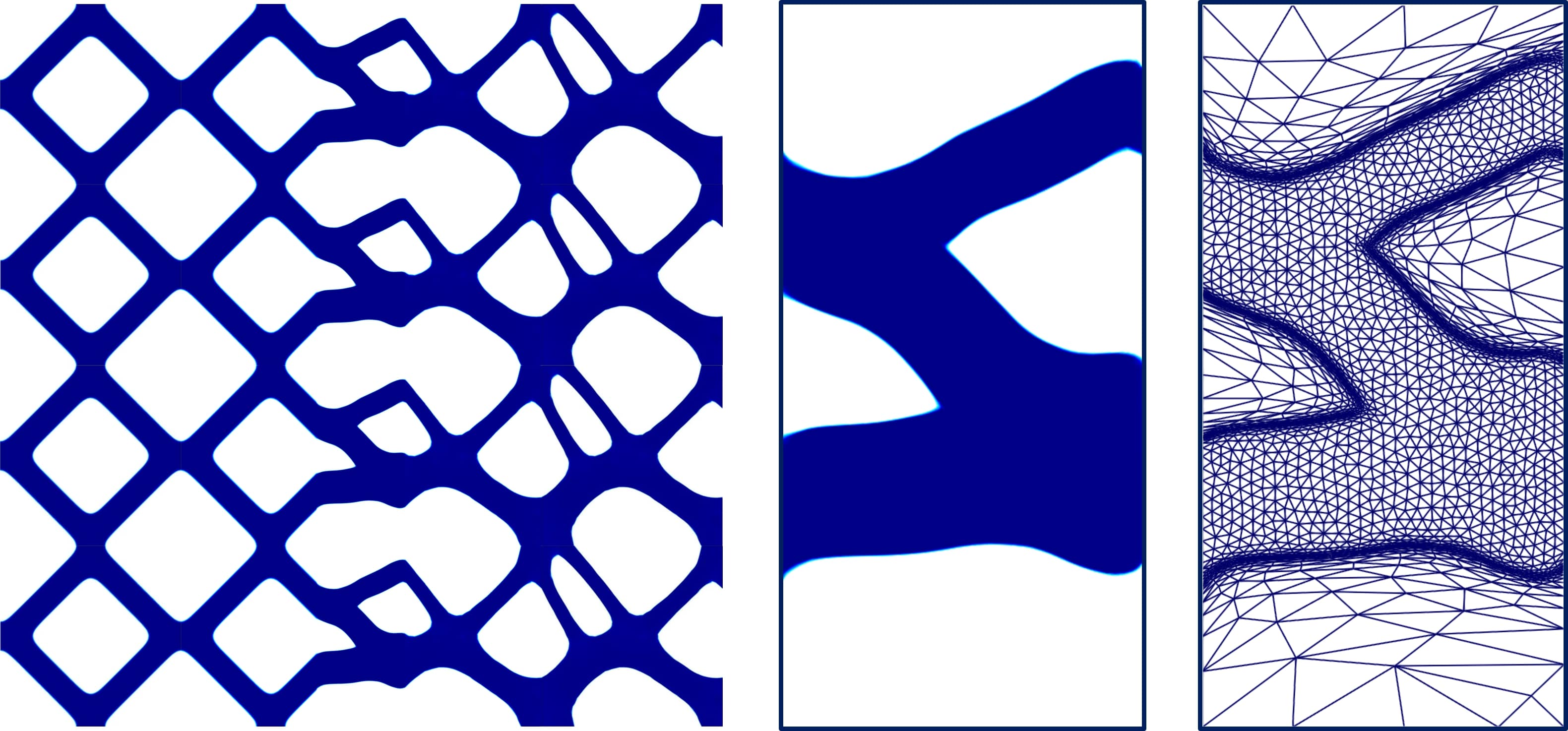}
	\caption{Matching A-B. $4 \times 4$ two-material ensemble (left); detail of the density distribution $\rho_Y$ and corresponding anisotropic adapted mesh (right) in $Y$.}
	\label{mesh}
\end{figure}

Figure~\ref{square_sim1} (bottom) displays the output of CONFLUENCE algorithm. In region $Y$ we can still recognize the initial topology of both the geometries, despite the complete absence of connectivity before morphing.
Figure~\ref{mesh} (left) shows the $4\times4$ two-material ensemble. The panel on the right highlights the smoothness characterizing the distribution of $\rho_Y$ together with the sharpness of the solid/void interface. This is the result of the anisotropic mesh adaptation process which selects the elements in an optimal way, in order to match the directional features of the design. We remark that an isotropic mesh is employed to discretize the internal part of the junction, according to the hybrid mesh generation paradigm described in Section~\ref{numerical_discretization}.

\begin{figure}[h!]
	\centering
	\includegraphics[width=0.55\textwidth]{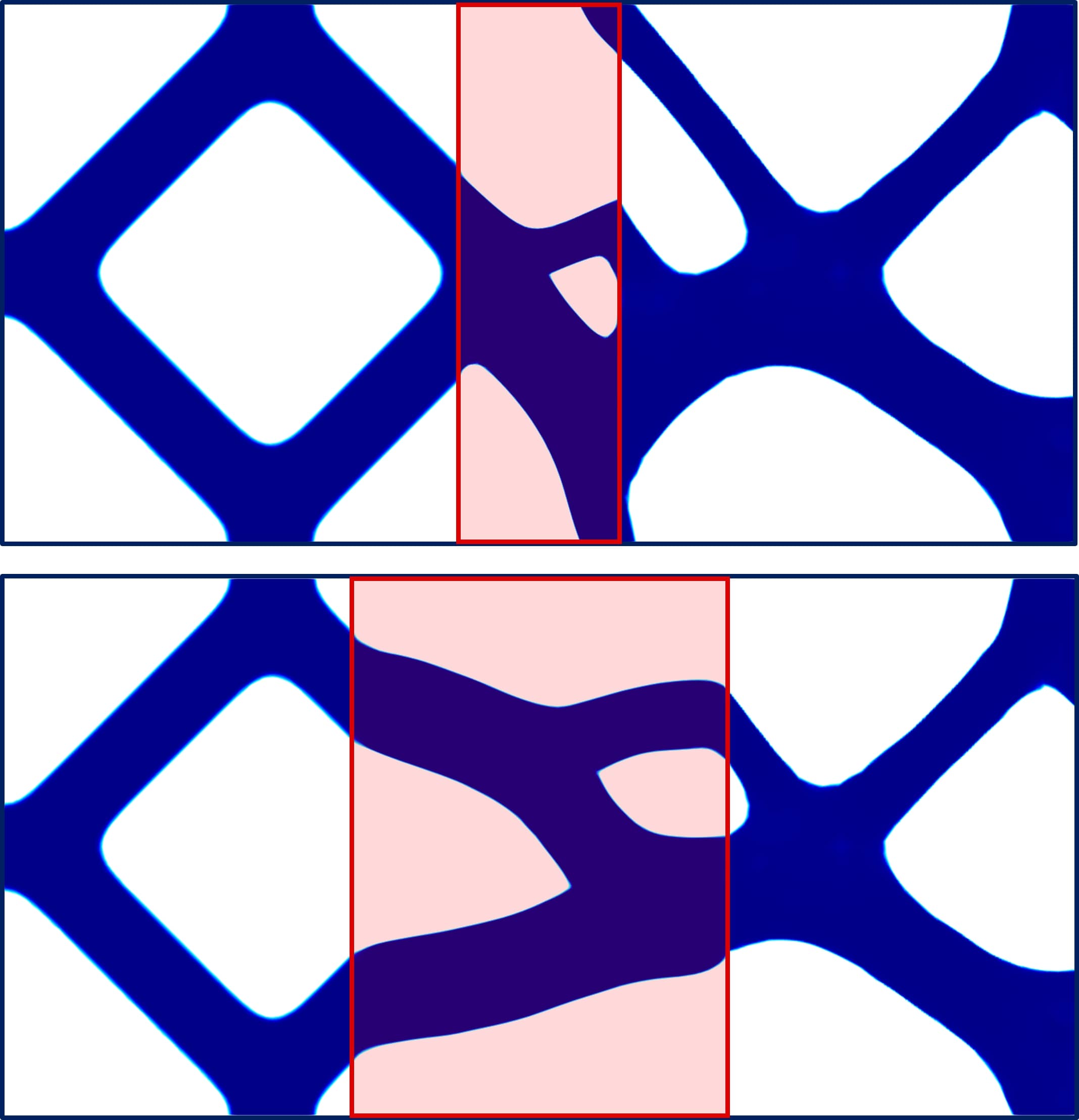}
	\caption{Matching A-B. Sensitivity of $\rho_{Y}$ to $\delta$: $\delta = 0.3$ (top) and $\delta = 0.7$ (bottom).}
	\label{geomAB_delta}
\end{figure}

We exploit this case study first to investigate the sensitivity of $\rho_Y$ to the value selected for $\delta$.
To this goal, we make two different choices for such parameter, namely $\delta = 0.3$ and $\delta = 0.7$, while preserving $s = 0$. In Figure~\ref{geomAB_delta}, we provide the corresponding results.
A cross-comparison among the three layouts in Figures~\ref{square_sim1} (bottom) and \ref{geomAB_delta} reveals that, for increasing values of the morphing region width, the designs miss the features characterizing the original geometries. Thus, for 
a too small value for $\delta$, we can lose some structural property of interest. For instance,
in the specific case of the top panel in Figure~\ref{geomAB_delta}, the right diagonal strut of geometry B is not modified by the morphing process so that the junction lacks horizontal connectivity. On the other hand, the connection along the vertical direction is reached only through the periodic replication of the density $\rho_{LYR}$.
On the contrary, the choice $\delta = 0.7$ leads to ignore the upward strut in cell B (bottom panel in Figure~\ref{geomAB_delta}) and to drastically change the original structural properties characterizing the two RVEs.

\begin{figure}[h]
	\centering
	\includegraphics[width=0.95\textwidth]{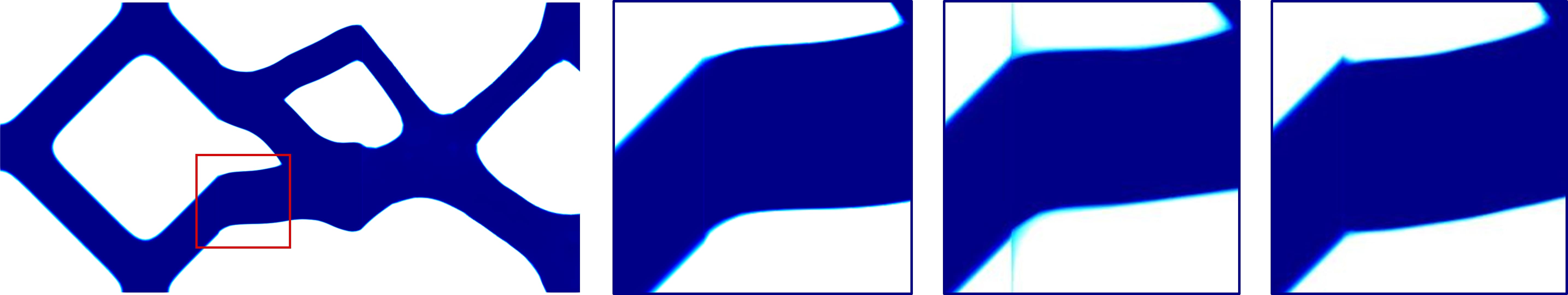}
	\caption{Matching A-B. Sensitivity of $\rho_Y$ to the matching conditions on the density and velocity: reference configuration (enlarged view on the left); $\rho_Y$ distribution when removing the matching of the density (center enlarged view) and of the velocity (enlarged view on the right).}
	\label{constraints_effect}
\end{figure}

On the same cell configuration, we analyze the different role played by the matching conditions on the density and on the velocity along $\Gamma_L$ and $\Gamma_R$. The enlarged view on the left in Figure~\ref{constraints_effect} provides the reference layout when both the matching conditions are applied. It is evident that such constraints ensure continuity as well as smoothness to the final density. On the contrary, when we neglect the requirements \eqref{min_topopt_discrete}$_3$ - \eqref{min_topopt_discrete}$_4$ on the final topology, spurious values for $\rho_Y$ arise, compromising the global continuity of $\rho_{LYR}$ (see the center enlarged view).
On the other hand, the removal of the matching condition on the velocity characterizing space $W_{{\bf g}, h}$ may yield sharp corners at the junction, thus degrading the global smoothness of the material in a neighbourhood of $\Gamma_L \cup \Gamma_R$. For instance, in the enlarged view on the right in Figure~\ref{constraints_effect}, we only impose the matching of the $x$-component of the velocities $\uu_h$ along $\Gamma_L$ and $\Gamma_R$.
\begin{figure}[h!]
	\centering
	\includegraphics[width=0.65\textwidth]{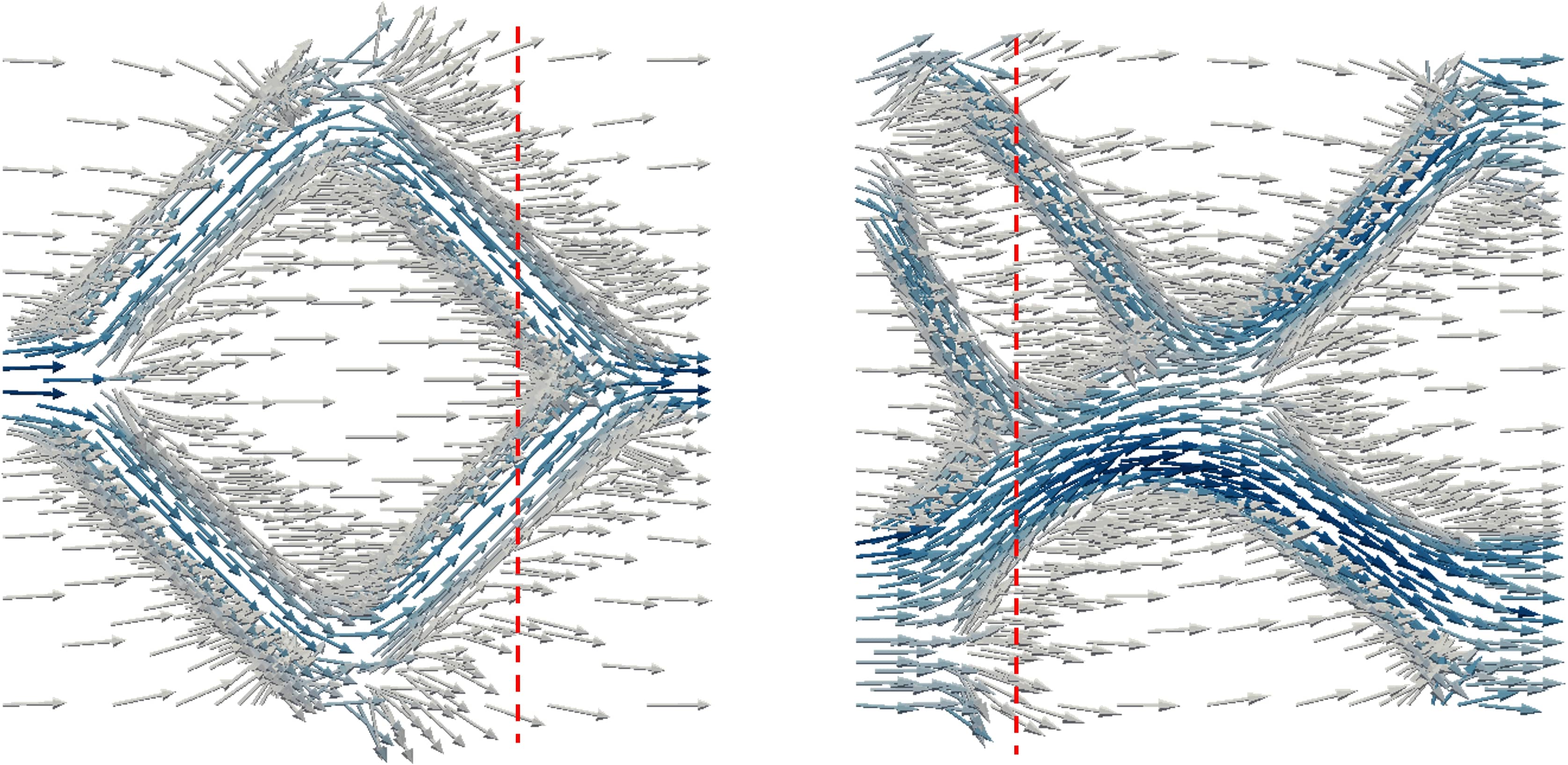}
	\caption{Matching A-B. Velocity fields $\uu_{h, L}$ in $Y_L$ (left) and $\uu_{h, R}$ in $Y_R$ (right), where $\Gamma_L$ and $\Gamma_R$ are red-highlighted.}
	\label{flux}
\end{figure}
This generates a small kink, which can be ascribed to the unconstrained behaviour allowed to the design procedure along the vertical direction (Figure~\ref{flux} highlights the relevance of both the components of fields $\uu_{h, L}$ and $\uu_{h, R}$ along $\Gamma_L$ and $\Gamma_R$).

\paragraph{Matching geometries B and D}

\begin{figure}[h]
	\centering
	\includegraphics[width=0.55\textwidth]{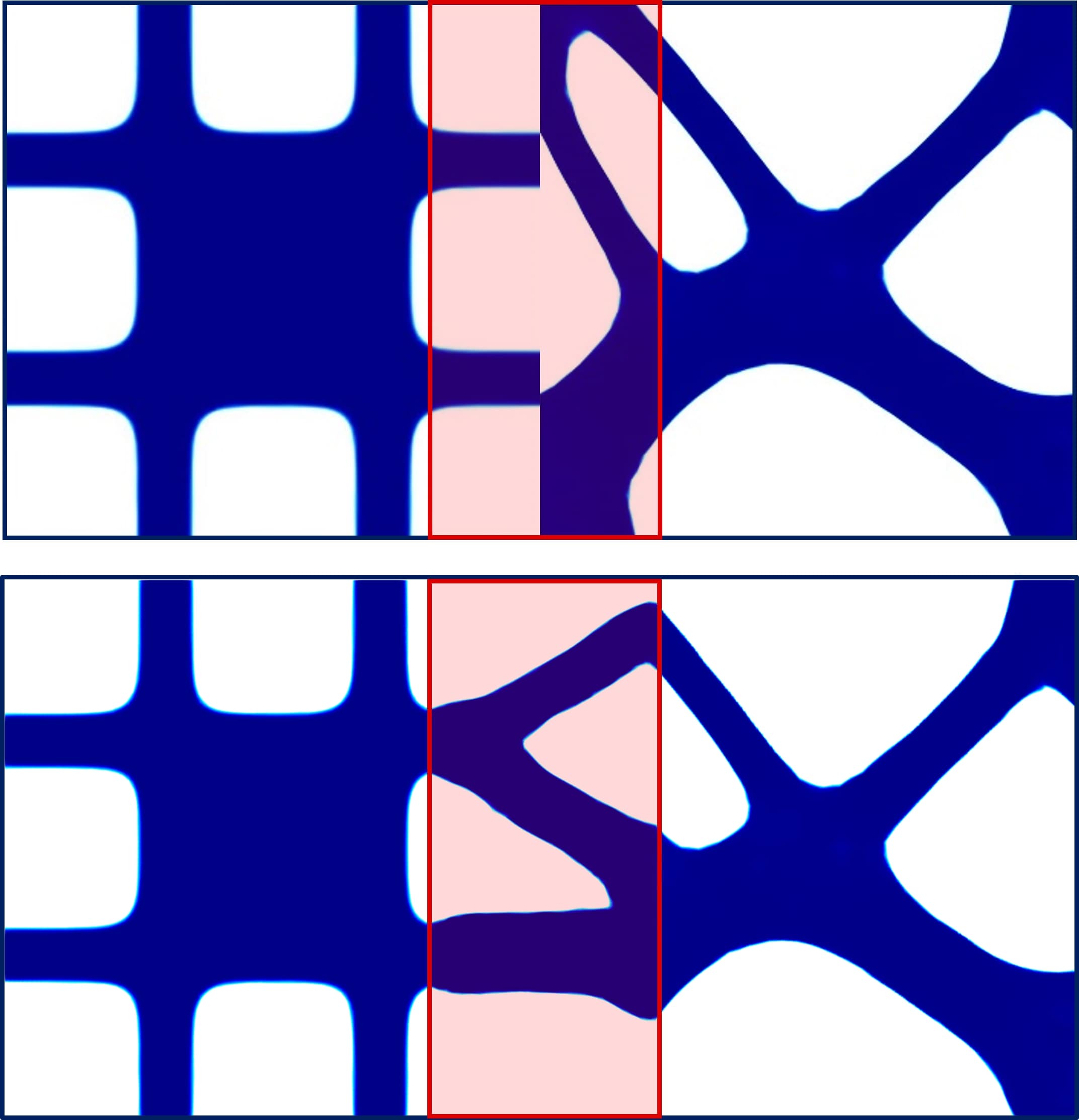}
	\caption{Matching B-D. Initial (top) and  final (bottom) density distribution provided by CONFLUENCE in $\Omega^\circ$.}
	\label{geomBD}
\end{figure}

We address the morphing from geometry B to geometry D in order to tackle the non-matching material at the common interface. 
The outcome of CONFLUENCE algorithm, for $\overline{\alpha} = 2.5 \, \mu \cdot 10^5$, $\delta = 0.43$ and $s = 0$, is shown in Figure~\ref{geomBD}. The effect of the morphing is to horizontally bend the diagonal struts of cell D in order to join the straight trusses in cell B.

\begin{figure}[h]
	\centering
	\includegraphics[width=0.65\textwidth]{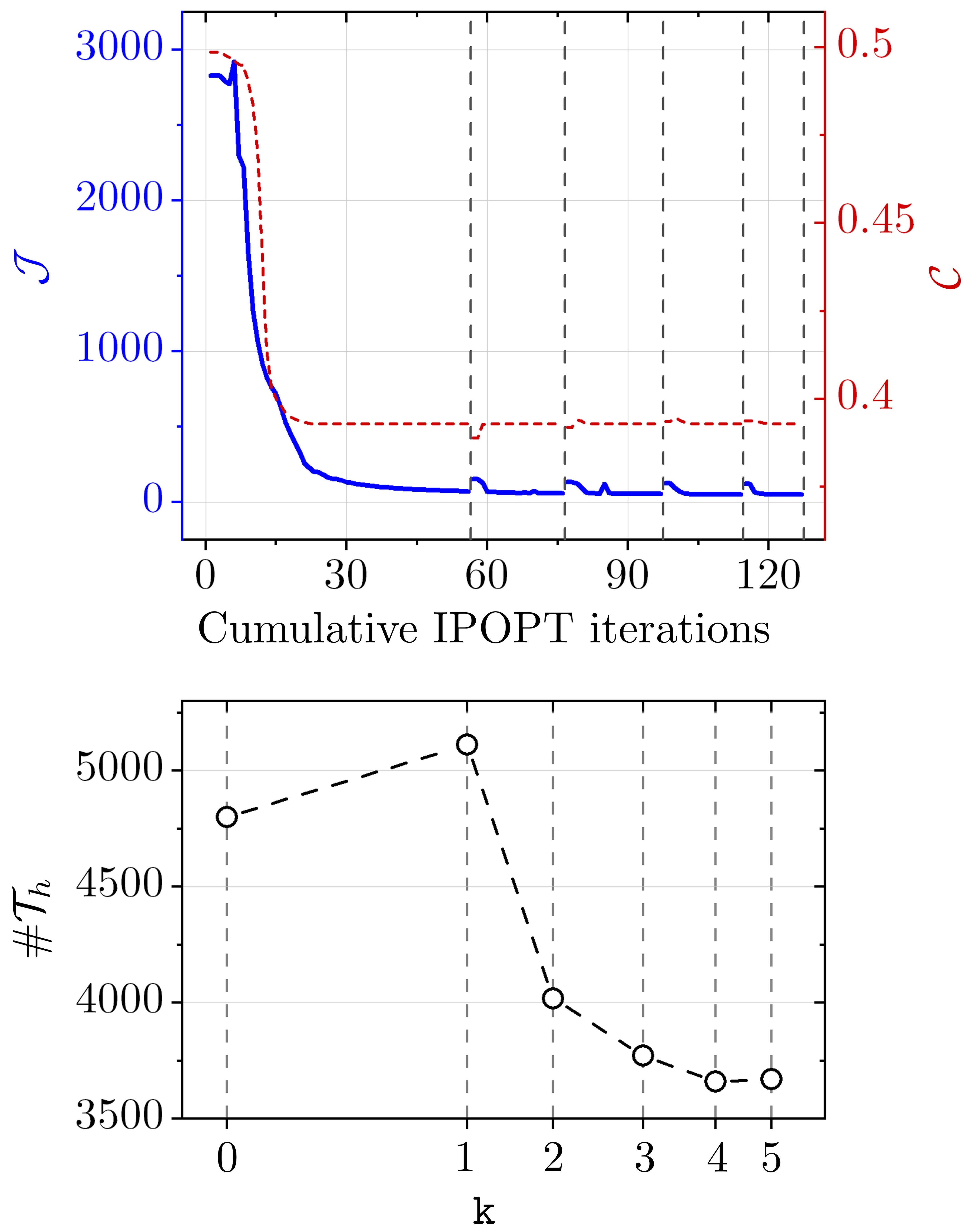}
	\caption{Matching B-D. Convergence history of the goal functional $\mathcal J$ and of the constraint $\mathcal C$ as a function of the cumulative IPOPT iterations (top);  mesh cardinality evolution across the adaptation process (bottom). The vertical dashed segments align the two panels in terms of the iteration index  $\tt k$.}
	\label{convergence}
\end{figure}

In this scenario, we focus on the performance of the optimization and on the evolution of the mesh elements throughout the adaptation process. To this aim, we plot the trend of $\mathcal{J}$ in \eqref{goal} and of the constraint $\mathcal{C} = \lvert Y \lvert^{-1} \int_Y \rho_h dY$ in \eqref{min_topopt_discrete}, as a function of the cumulative number $\sum_{\tt k} \xi_{\tt k}$ of IPOPT iterations (see Figure~\ref{convergence} (top)). Both the functional and the constraint quickly converge, except for mild oscillations. A different behaviour characterizes the cardinality, $\# \mathcal{T}_h$, of the computational mesh as a function of index $\tt k$. Figure~\ref{convergence} (bottom) exhibits the evolution typical of a mesh adaptation process, which includes an initial abrupt increment of the cardinality, followed by a gradual reduction of the number of triangles until stagnation. The plots in Figure~\ref{convergence} highlight the twofold control performed by the tolerances $\tt TOPT$ and $\tt CTOL$.
Indeed, the tolerance $\tt TOPT$ is guaranteed already on the first adapted mesh. On the contrary, the stagnation of the relative mesh cardinality within tolerance $\tt CTOL$, is ensured only after $5$ adaptation steps (i.e., $5$ complete while loops). The outcome of this coupled check guarantees an optimized solution both from a structural and a computational viewpoint.

\begin{figure}[h]
	\centering
	\includegraphics[width=0.55\textwidth]{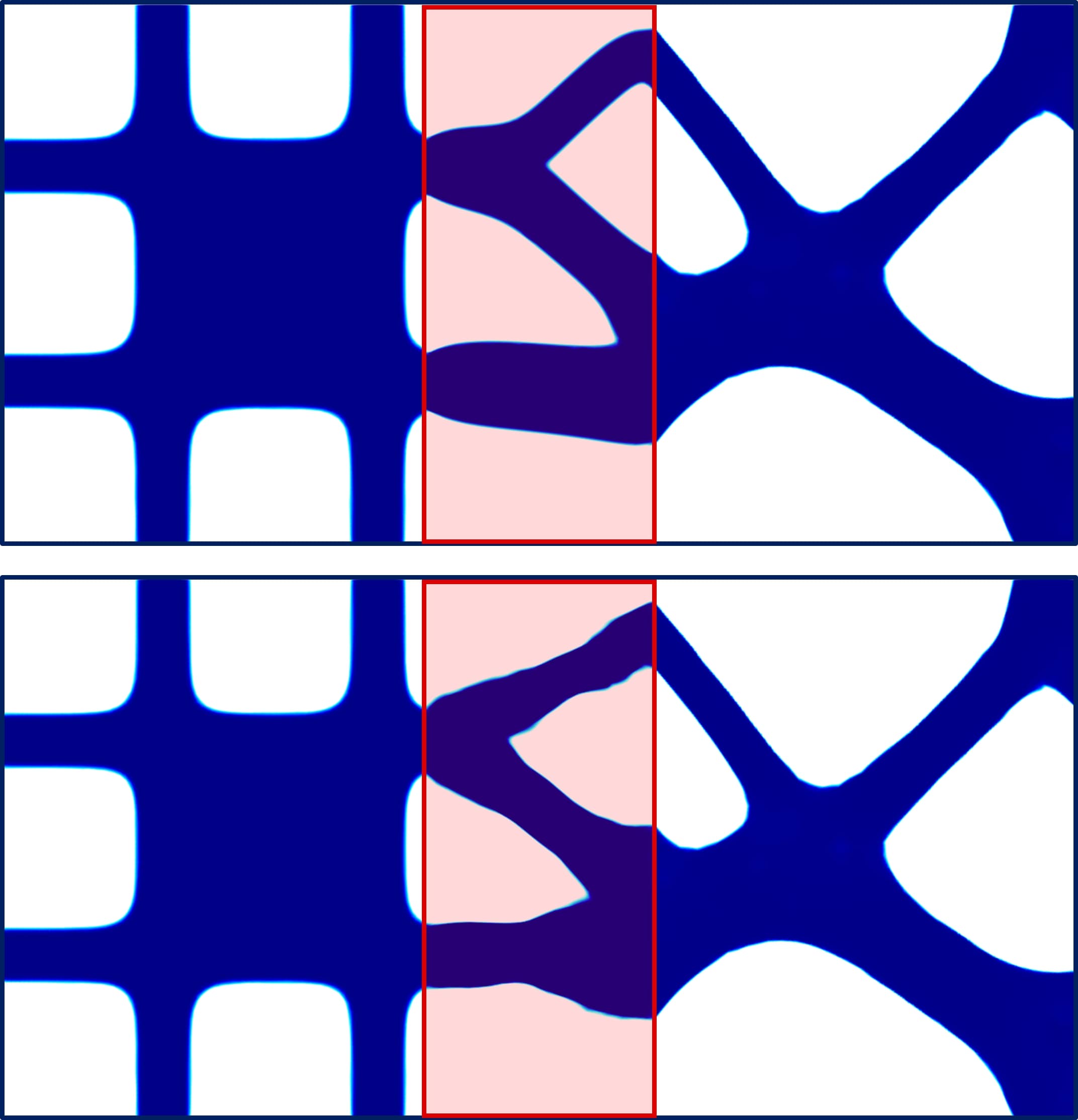}
	\caption{Matching B-D. Sensivity of $\rho_Y$ to the upper bound $\overline{\alpha}$ for the inverse permeability $\alpha$: $\overline{\alpha} = 2.5 \, \mu \cdot 10^4$ (top), $\overline{\alpha} = 2.5 \, \mu \cdot 10^6$ (bottom).}
	\label{grid_sim1}
\end{figure}

On this configuration, we carry out a sensitivity analysis of the CONFLUENCE output with respect to the value for $\overline{\alpha}$ in \eqref{alpha_range}. In particular, we adopt a larger and a smaller value for $\overline{\alpha}$ with respect to the one considered in Figure~\ref{geomBD}. In Figure~\ref{grid_sim1}, we provide the distribution of $\rho_{LYR}$ for $\overline{\alpha} = 2.5 \, \mu \cdot 10^4$ (top) and $\overline{\alpha} = 2.5 \, \mu \cdot 10^6$ (bottom). 
\\
We do not detect a strong dependence of the final topology on parameter $\overline{\alpha}$. However, a reduced permeability (i.e., a larger value for $\overline{\alpha}$) seems to hinder the Stokes flow-driven topology design, as confirmed by the wavy contour of the optimized density in the junction area (compare top and bottom panels in Figure~\ref{grid_sim1}).

\paragraph{Matching geometries B and C}

\begin{figure}[h]
	\centering
	\includegraphics[width=0.55\textwidth]{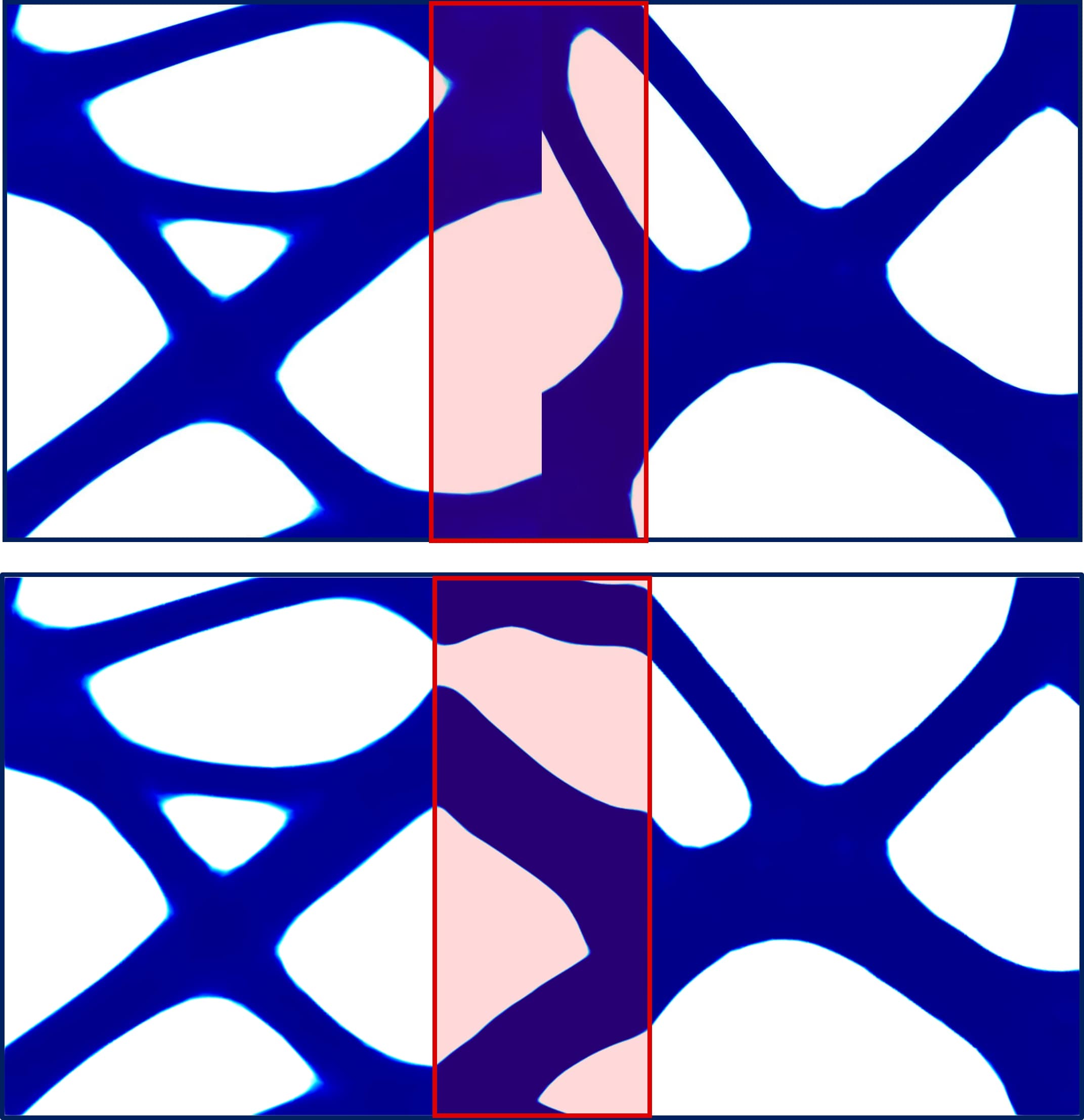}
	\caption{Matching B-C. Initial (top) and  final (bottom) density distribution provided by CONFLUENCE in $\Omega^\circ$.}
	\label{geom_BCs0}
\end{figure}

The last matching that we consider combines two unit cells, which share a portion of material along the common interface $\mathcal E$, namely geometry B (on the left) and geometry C (on the right). For this purpose, we set $\overline{\alpha} = 2.5 \, \mu \cdot 10^5$, $s = 0$ and $\delta = 0.4$. We show the result of the morphing associated with such configuration in Figure~\ref{geom_BCs0}. The topology identified in $\Omega^\circ$ is characterized by long, hanging, horizontal struts, which can be sub-optimal in view of a mechanical analysis. 

\begin{figure}[h]
	\centering
	\includegraphics[width=0.55\textwidth]{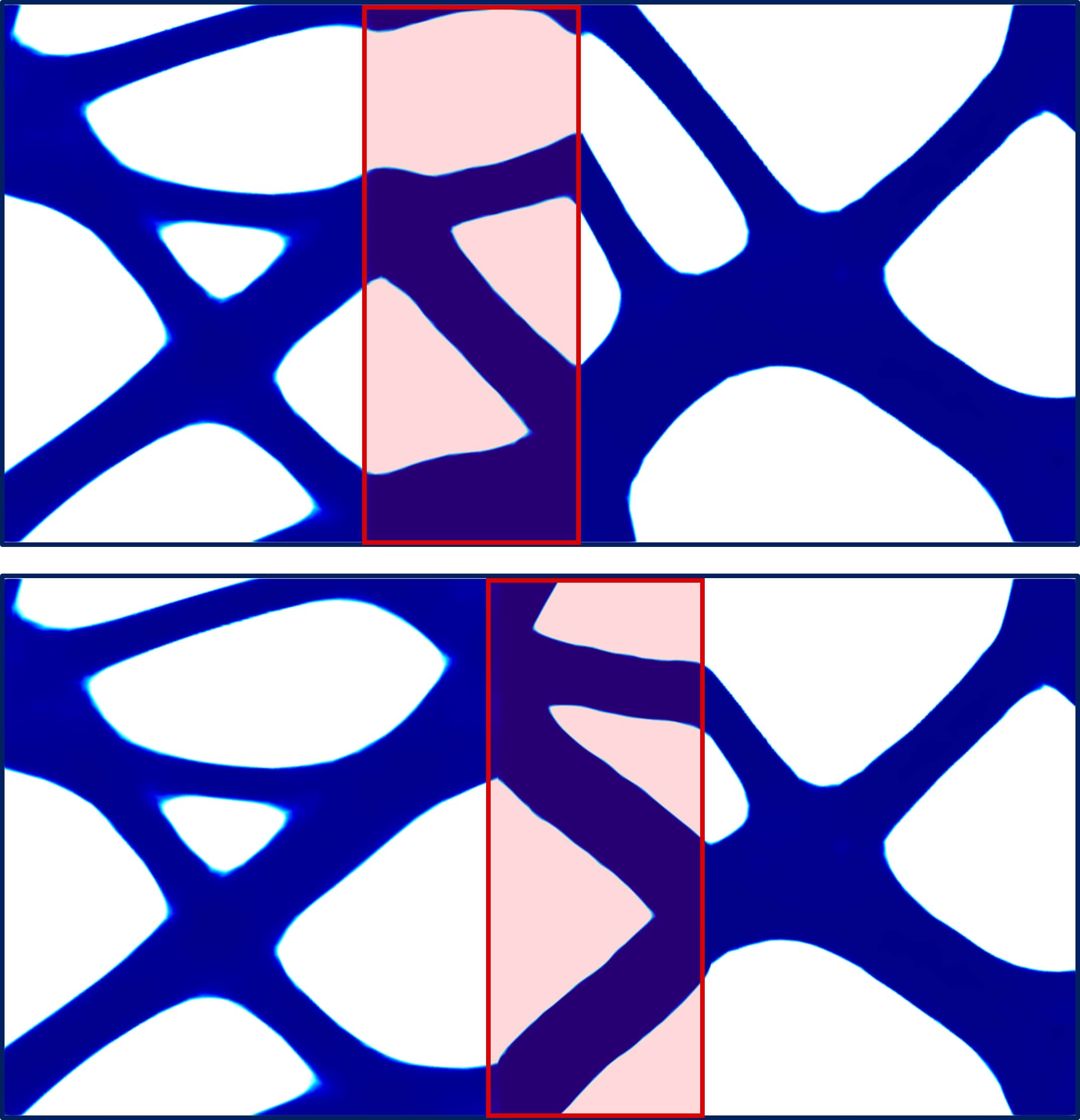}
	\caption{Matching B-C. Sensitivity of $\rho_Y$ to $s$: $s = -0.12$ (top), $s = 0.12$ (bottom).}
	\label{sim2_sim1}
\end{figure}

These two geometries are adopted to evaluate the impact of the shift parameter $s$ onto the output of CONFLUENCE algorithm. This investigation is performed by setting in the previous configuration the values $s = -0.12$ and $s = 0.12$. Figure~\ref{sim2_sim1} collects the results of this analysis. The position of the interfaces $\Gamma_L$ and $\Gamma_R$ turns out to be crucial in order to guarantee the generation of topologies suited for practical applications. As a matter of fact, the choice $s = -0.12$ (see Figure~\ref{sim2_sim1} (top)) suffers from the similar drawbacks as the design in Figure~\ref{geom_BCs0} (i.e., the presence of vertically unsopported trusses). 
Conversely, for $s = 0.12$ (Figure \ref{sim2_sim1} (bottom)),
the two cells are well-connected, without any artifacts which may deteriorate the mechanical performance. This considerations support the choice of this last scenario as the best configuration for the mechanical analysis in Section~\ref{structural_analyses}.


\subsection{Generalization to a non-Cartesian morphing region}
\begin{figure}[h!]
	\centering
	\includegraphics[width=0.65\textwidth]{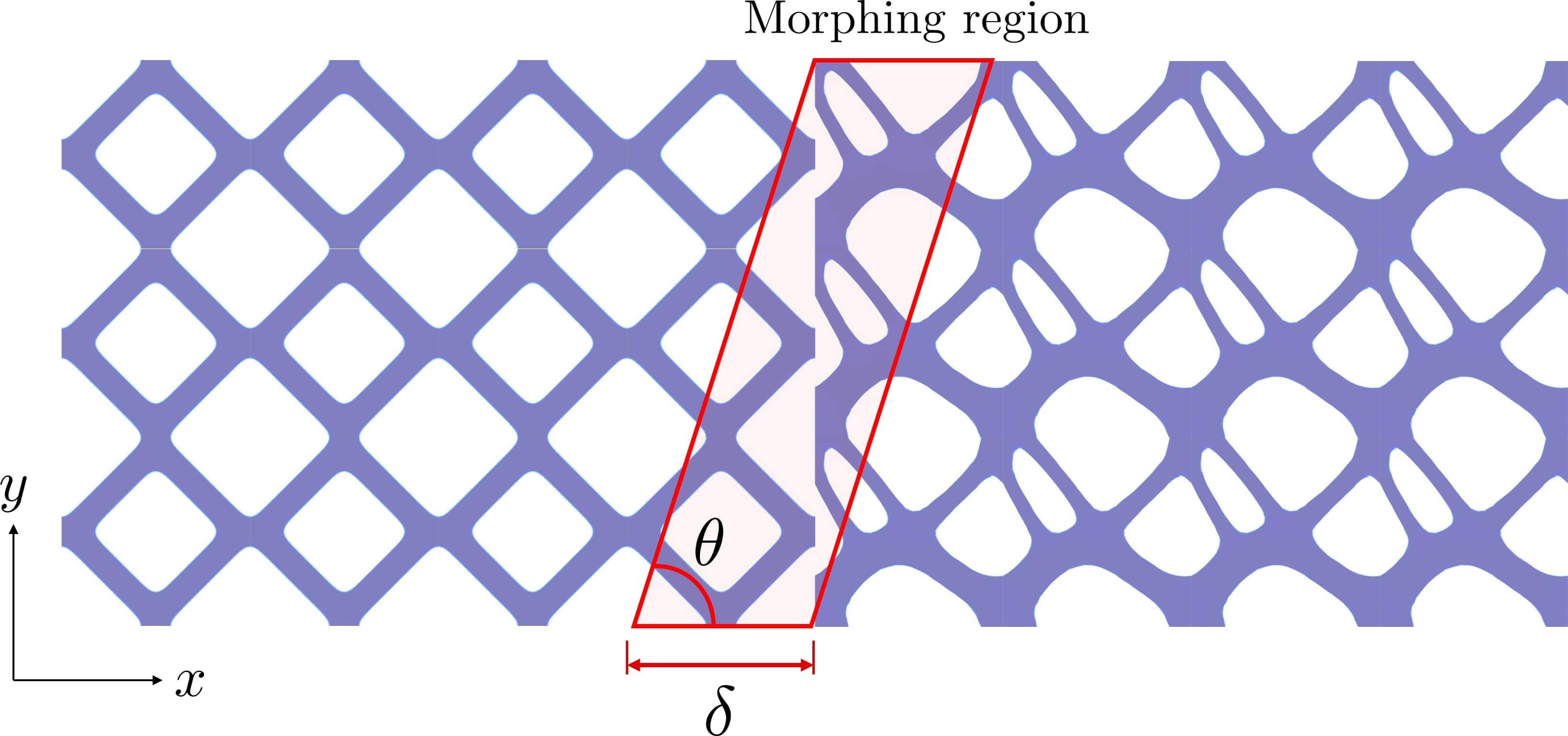}
	\caption{Non-Cartesian morphing region. Definition of the main geometric parameters.}
	\label{morphing_inclined}
\end{figure}

Practical applications often demand to join different cellular materials along an interface which is not vertical.
In this section, we consider a possible generalization of Algorithm \ref{algo} in order to tackle settings such as the one in Figure~\ref{morphing_inclined}. Here, the morphing region $Y$ coincides with a parallelogram which is overlapped to the lattices associated with different unit cells. In particular, $Y$ is characterized by an inclination $\theta$ with respect to the $x$-axis and by the width $\delta$. The main change to be done in CONFLUENCE algorithm consists in replacing the periodic boundary conditions along $\partial Y \setminus (\Gamma_L \cup \Gamma_R)$ with a homogeneous Neumann data.

\begin{figure}[h]
	\centering
	\includegraphics[width=0.55\textwidth]{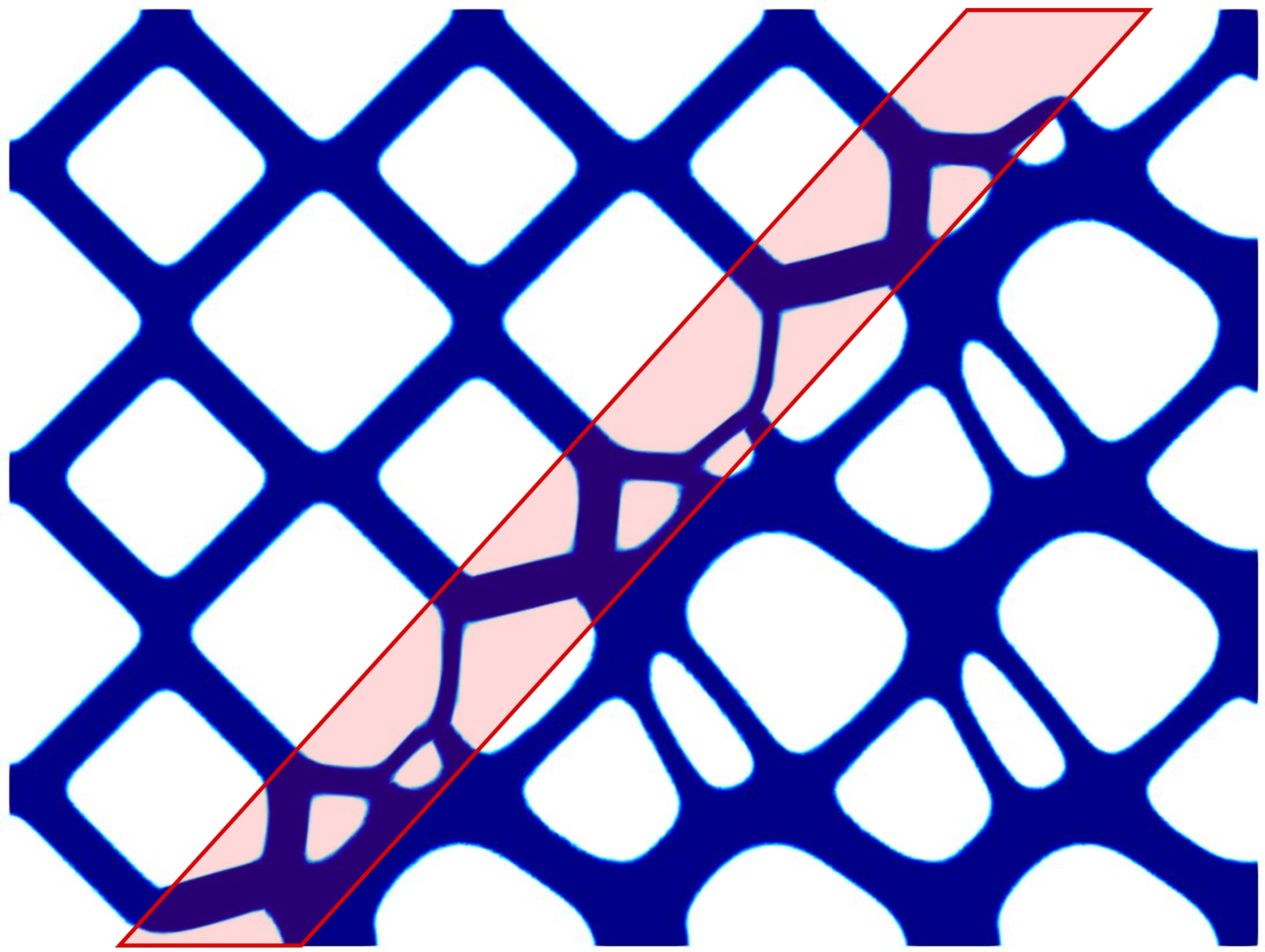}
	\\[1mm]
	\includegraphics[width=0.55\textwidth]{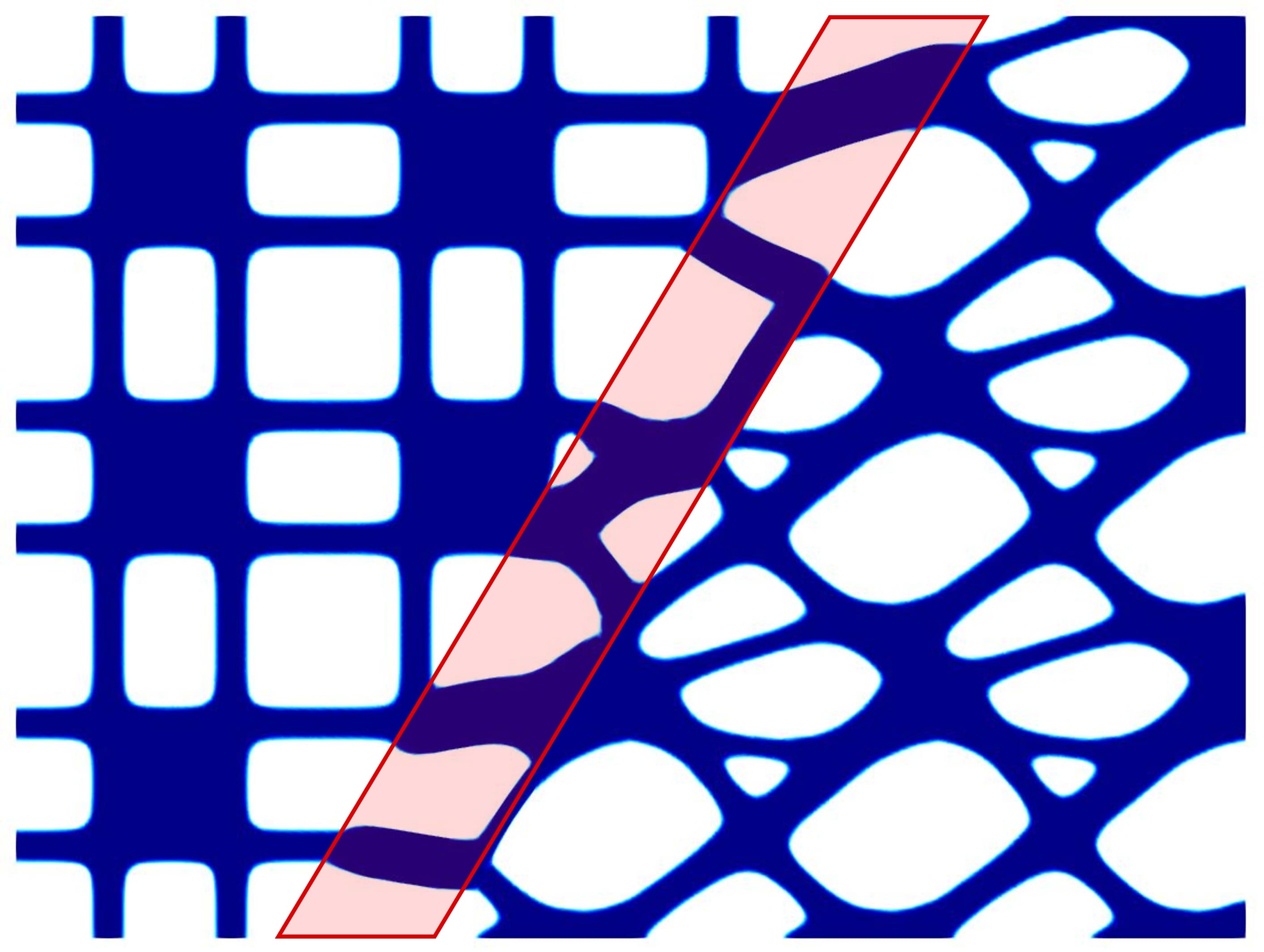}
	\caption{Non-Cartesian morphing region. Matching A-B for $\theta = \pi/4$ (top); matching D-C for $\theta = \pi/3$ (bottom). }
	\label{inclined_pi3_pi4}
\end{figure}

As benchmark configurations, we perform two matchings by morphing the $4\times 3$-lattice materials associated with geometries A (left) and B (right), and with geometries D (left) and C (right). For both these choices, we set $\overline{\alpha} = 2.5 \, \mu \cdot 10^6$ and $\delta = 0.4$, while the inclination $\theta$ is equal to $\pi/4$ for the first case (A-B) and to $\pi/3$ for the second case (D-C).
\\
Figure~\ref{inclined_pi3_pi4} gathers the resulting materials. In particular, we remark the extremely localized effect of the morphing, together with the absence of unsupported struts along both the $x$ and $y$-directions.

This preliminary assessment confirms the high flexibility of CONFLUENCE algorithm to tackle configurations which are still rarely addressed in the reference literature.

\section{Structural analysis}\label{structural_analyses}

\begin{figure}[h!]
\centering
	\includegraphics[width=0.55\textwidth]{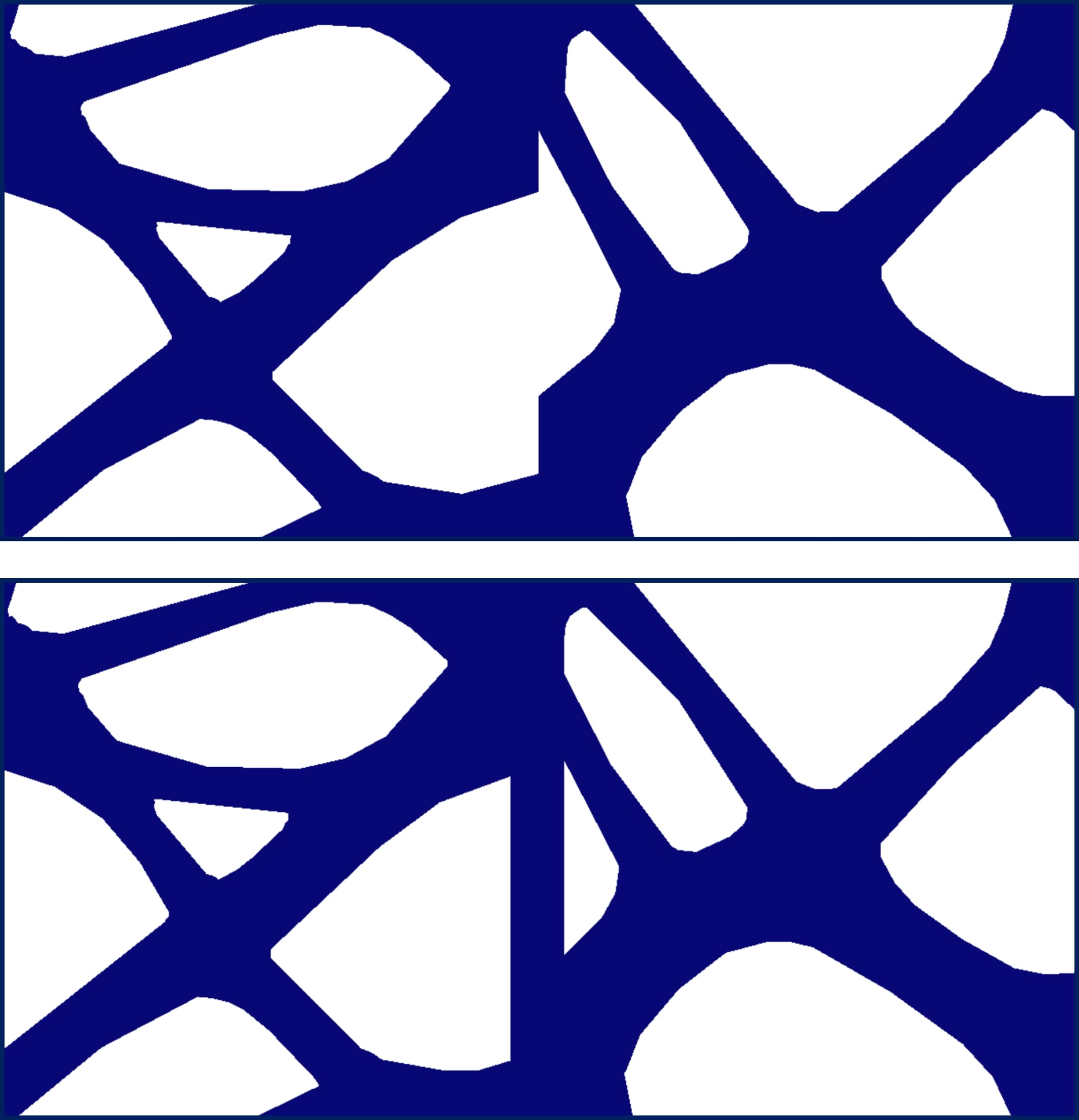}
	\caption{Structural analysis. Side-by-side (top) and solid wall (bottom) connection between cells B and C.}
	\label{conn_opts}
\end{figure}

The mechanical performance of the 
morphing geometry B-C in Figure~\ref{sim2_sim1} (bottom) is assessed by means of a dedicated finite element mechanical analysis. To this aim, we investigate the global response of the heterogeneous structure to a given load and we quantify possible stress localizations induced by the change of topology in the morphing region. Moreover, to further validate the effectiveness of CONFLUENCE algorithm, we carry out a comparison of the geometry B-C after morphing with the case of a straightforward side-by-side connection between cells B and C (see Figure~\ref{conn_opts} (top)), and with 
a connection represented by a solid wall of thickness $0.1$ separating the two materials (see Figure~\ref{conn_opts} (bottom)). We remark that while the side-by-side solution is not always necessarily pursuable (see Figures~\ref{square_sim1} and~\ref{geomBD} (top), where the cells do not share portions of material along the interface $\mathcal E$), the solid wall approach unavoidably increases the global mass of the multicellular specimen.
\begin{figure*}[t!]
	\centering
	\includegraphics[width=0.99\textwidth]{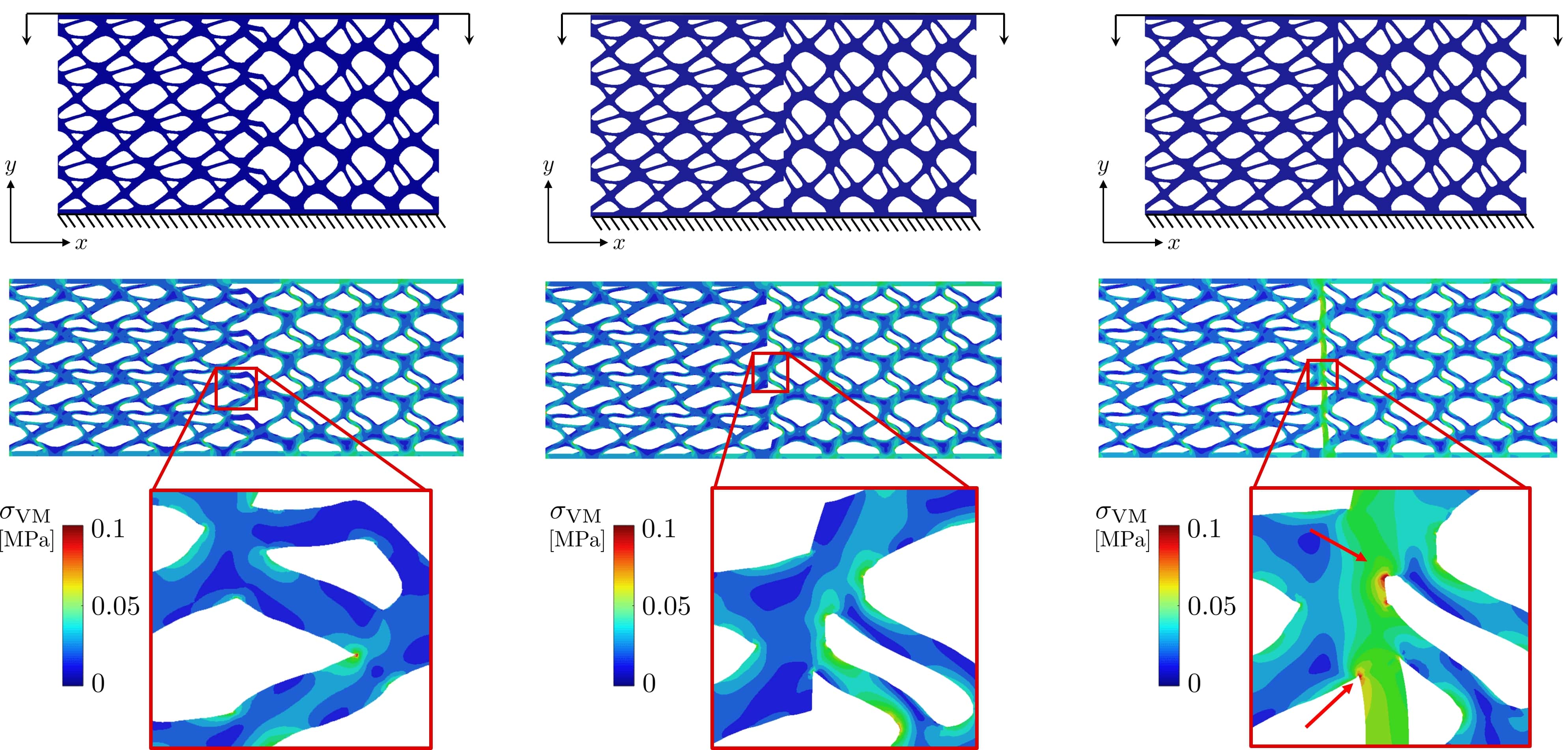}
	\caption{Structural analysis. Composite structures B-C for three matching strategies under a compressive displacement: CONFLUENCE algorithm (left), side-by-side connection (center), solid wall connection (right). Layouts and case study setting (top), von Mises stress distribution on the deformed structures with enlarged views (bottom).}
	\label{comp_disp}
\end{figure*}

The structural analysis is performed by using the commercial finite element software Abaqus\footnote{Abaqus, Dassault Syst{\`e}mes Simulia Corp, United States.}.
In Figure~\ref{comp_disp} (top), we provide the three settings to be compared, which consist of a heterogeneous volume composed by a $4 \times 4$-block of unit cells B joint with a $4 \times 4$-block of unit cells C. We apply a uniform vertical compressive displacement on the top of the multilattice material; we constrain the bottom along the $y$-direction while allowing for lateral expansion. Concerning the boundary configuration, two solid layers with thickness $0.1$ are introduced at the top and at the bottom in order to mimic a sandwich structure. Finally, the displacement along the vertical border is left free.\\
The considered base material is characterized by a linear elastic behaviour with unitary Young's modulus and a Poisson's ratio equal to $0.3$.
\\
A quadratic finite element approximation is used to discretize the structural displacement.

The response of the composite structures to the applied displacement is investigated in Figure~\ref{comp_disp} (bottom), where we show the distribution of the von Mises stress $\sigma_{\textrm{VM}}$ on the deformed multilattice specimen, together with enlarged views in correspondence with the junction.
The deformation and the von Mises stress distribution of the three materials are essentially identical far from the morphing region. On the contrary, the choice adopted to join the different lattices leads to a significant difference, in particular on the stress, in correspondence with the area around the interface $\mathcal E$. We observe an overall increment for the stress in both the non-optimized configurations (compare the three detailed views), together with a significant stress localization for the solid wall solution.

We compare the three different scenarios in Figure~\ref{comp_disp} also in terms of the pressure measured along the $x$-direction at the top border where the displacement is applied. As expected, CONFLUENCE algorithm and the side-by-side strategy exhibit a global similar trend, whereas the solid wall solution leads to a high pressure peak in correspondence with the connection. This anomalous feature might represent an issue, in particular when a smooth transition of the mechanical properties among the different materials is demanded.

\begin{figure*}[t!]
	\centering
	\includegraphics[width=0.99\textwidth]{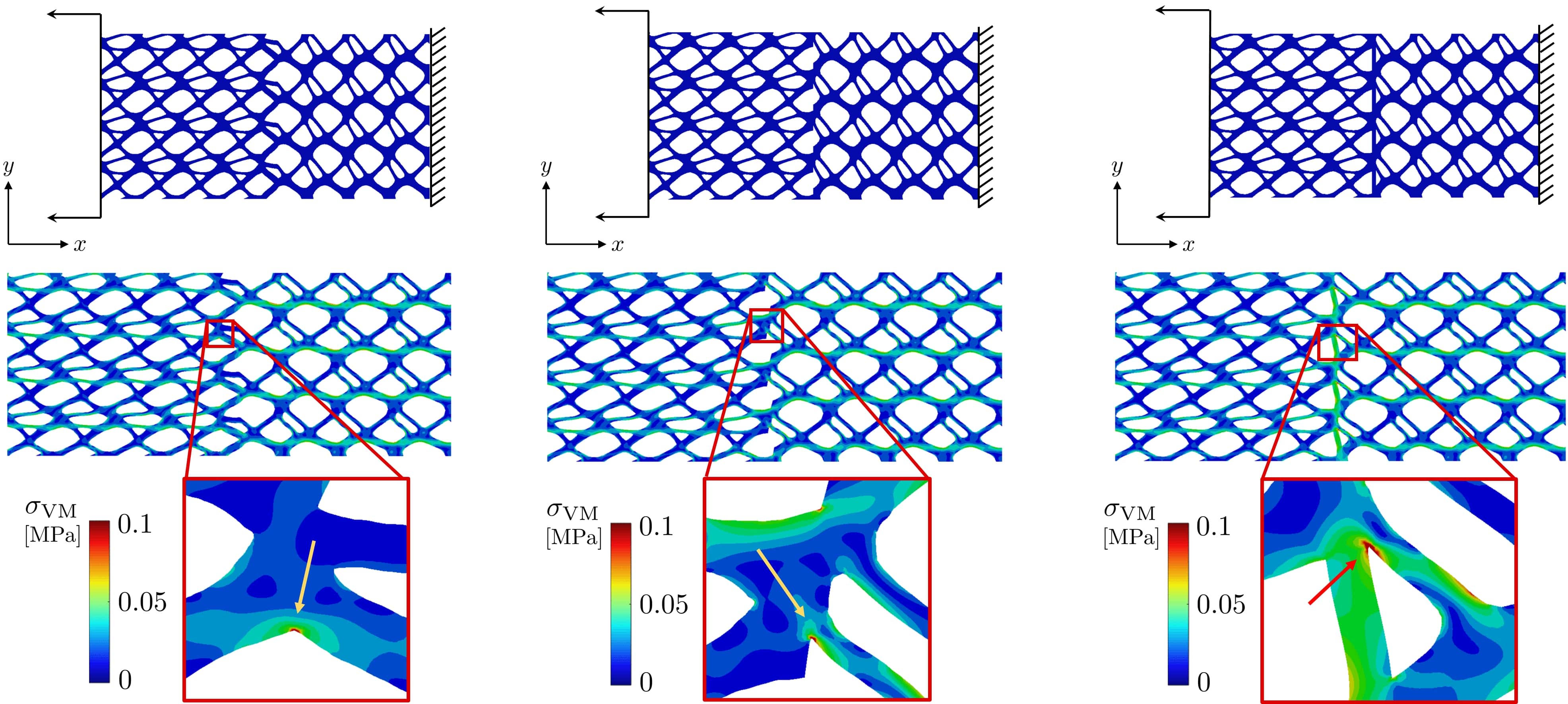}
	\caption{Structural analysis. Composite structures B-C for three matching strategies under a tensile displacement: CONFLUENCE algorithm (left), side-by-side connection (center), solid wall connection (right). Layouts and case study setting (top), von Mises stress distribution on the deformed structures with enlarged views (bottom).}
	\label{traz_disp}
\end{figure*}

We replicate the analysis above by changing the case study setting. A tensile displacement is now applied along the left side, while the composite material is clamped along the right border. The top and the bottom of the specimen are not subject to any imposed displacement.
\\
Figure~\ref{traz_disp} collects the results of this new framework by providing the same information as in Figure~\ref{comp_disp}. It turns out that this setting is in general more challenging with respect to the compressive case, as the von Mises distribution highlights. CONFLUENCE algorithm delivers the most effective solution when compared with the side-by-side and the solid wall connections. In particular, we notice that the stress localization is much more severe when resorting to the solid wall approach.
\begin{figure}[h!]
\centering
	\includegraphics[width=0.55\textwidth]{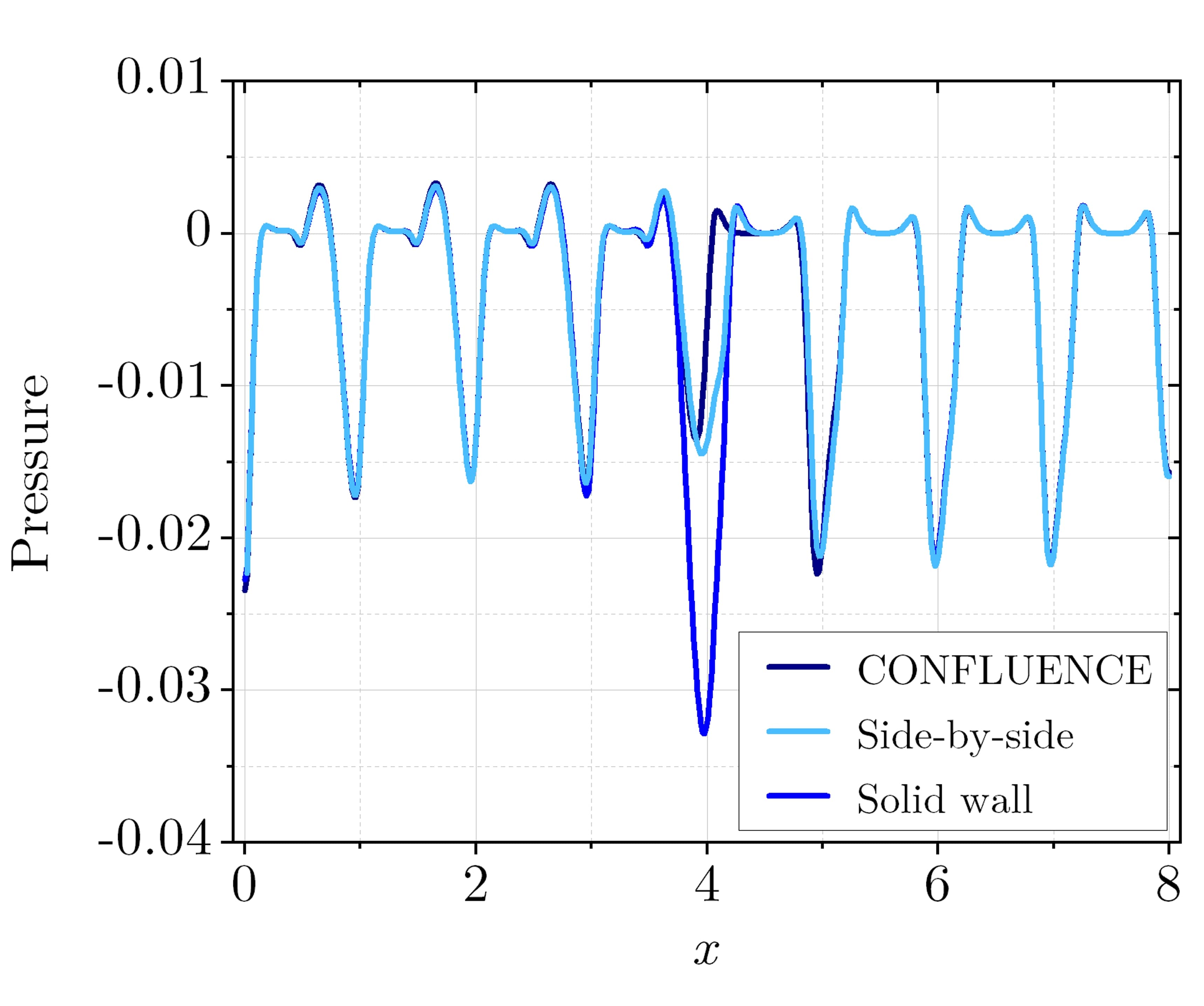}
	\caption{Structural analysis. Trend of the pressure as a function of $x$ along the top border of the specimen.}
	\label{comp_pressure}
\end{figure}
%

\begin{figure}[h!]
\centering
	\includegraphics[width=0.6\textwidth]{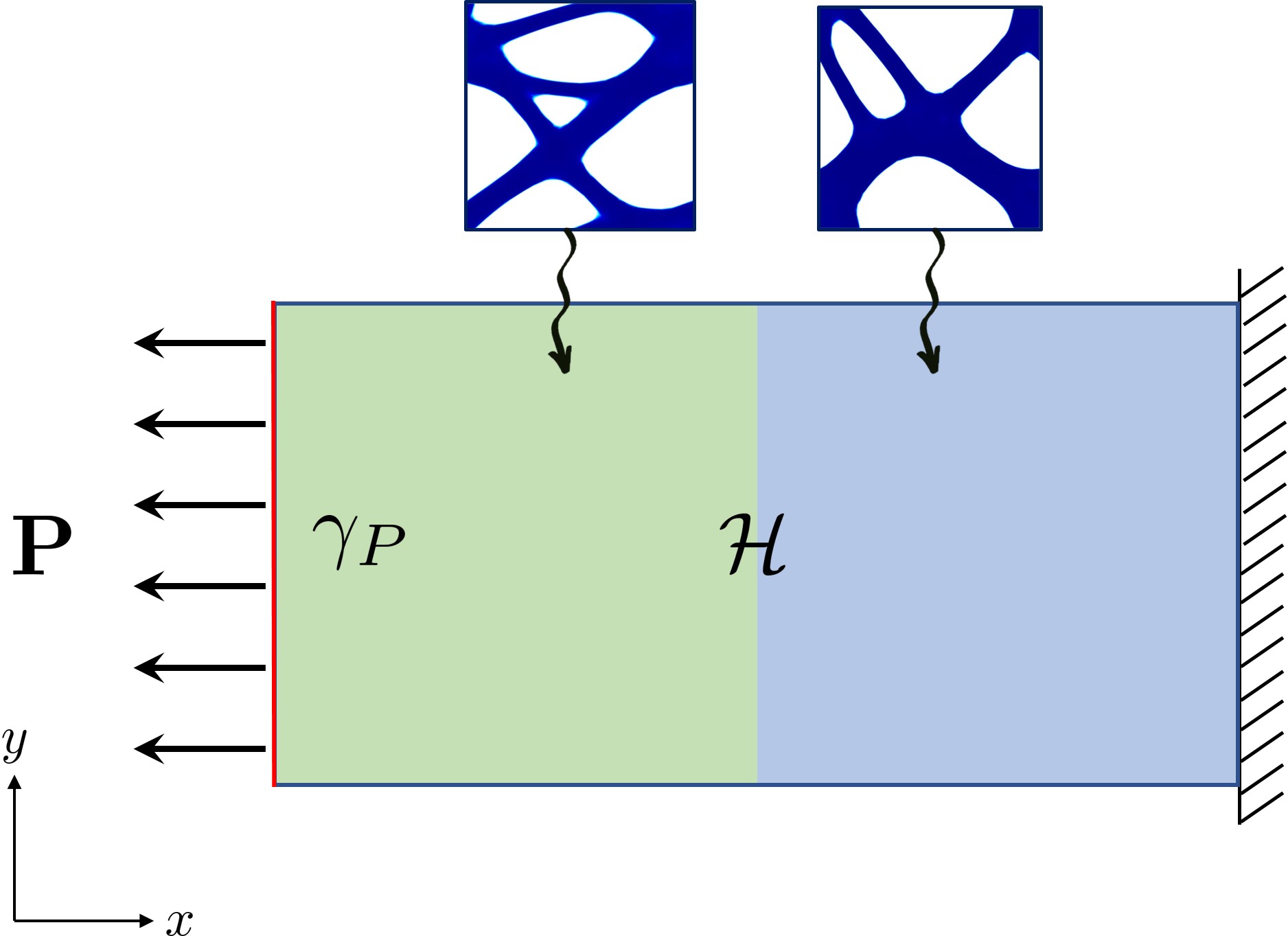}
	\caption{Structural analysis. Sketch of the homogenized specimen $\mathcal H$: material and mechanical setting.}
	\label{sketch_homogenized}
\end{figure}

As a final check, we consider 
a specimen, $\mathcal H$, of the same global extension as the three composite structures in Figure~\ref{traz_disp}, characterized by discontinuous mechanical features. In particular, as sketched in Figure~\ref{sketch_homogenized}, the left and the right halves of the material share the homogenized properties associated with cell B and C, respectively.\\ 
We compute the compliance, $C^* =  \int_{\gamma_P} {\bf P} \cdot {\bf d}\, ds$, associated with the load $\bf P$ applied to the boundary portion ${\gamma_P}\subset \partial \mathcal H$ and inducing the displacement $\bf d$ (see Figure~\ref{sketch_homogenized}). 
The same mechanical configuration is enforced on the multilattice materials in Figure~\ref{traz_disp}, for comparison purposes. 
\\
Table~\ref{tab:tab1} gathers the results of such an analysis in terms of compliance and percentage error with respect to the homogenized configuration $\mathcal H$. As expected, the solid wall connection provides the smallest compliance error while representing a non viable strategy in practice (see Figures~\ref{comp_disp}-\ref{traz_disp}). CONFLUENCE connection turns out to better perform with respect to the side-by-side solution, leading to a lower compliance and to half of the percentage error, approximately.
\begin{table}[h]
    \centering
    \begin{tabular}{ccc}
    \hline
                       & $C^*$ & $E_\%$ \\
                       \hline
         ${\mathcal H}$ & 19.12 &       \\[1mm]
         morphing      & 21.78 & 13.88  \\[1mm]
         side-by-side  & 23.80 & 24.98  \\[1mm]
         solid wall    & 19.88 &  3.95  \\
         \hline\\
    \end{tabular}
    \caption{Structural analysis. Quantitative comparison in terms of compliance and percentage error between the homogenized specimen $\mathcal H$ and CONFLUENCE, side-by-side and solid wall connections.}
    \label{tab:tab1}
\end{table}

\section{Conclusions}\label{concl}
We propose the new algorithm CONFLUENCE (CONnection by FLUids of differENt CElls) to design the junction between heterogeneous lattice materials. To this end, we exploit a Stokes-type topology optimization setting which is used to locally modify the matching between different RVEs. 

CONFLUENCE guarantees a cost-effective and local re-design of the materials, thus inducing a very mild modification to the original behaviour of the matched materials.

The new morphing process is assessed by investigating the sensitivity of the final optimized layout to different geometric and physics parameters in order to validate the robustness of the algorithm.
Additionally, CONFLUENCE algorithm is compared with two basic approaches (side-by-side and solid wall connections) used to join different materials, in terms of displacement and von Mises stress. As far as the configuration here considered (composite structure B-C), the matching optimized through CONFLUENCE exhibits good properties. The morphing acts on a very small area so that the mechanical performance of the joint RVEs is slightly affected. Moreover, the solution offered by the CONFLUENCE avoids undesired stress concentration and limits any pressure peak in correspondence with the morphing region.

Finally, the new algorithm is challenged with a non-Cartesian-aligned transition region to mimic practical configurations where the morphing between two consecutive heterogeneous lattices takes place along a generic curve. Results in Figure~\ref{inclined_pi3_pi4}, yet preliminary, confirm the robustness of CONFLUENCE also in such a context.

The present work opens some relevant issues for future investigations, such as:
\begin{itemize}
    \item the extension of the presented algorithm to a $3$D setting with a view to engineering applications;
    \item the inclusion in the optimization problem \eqref{min_topopt} of additional structural constraints (e.g., stiffness, stress, buckling), to take into account, for instance, the conclusions drawn in Section~\ref{structural_analyses};
    \item the selection of a driver for the topology optimization in the morphing region different from the Stokes problem in Section~\ref{method} (see, e.g.,~\cite{li2016});
    \item the generalization of the present setting to a parametric framework in order to exploit consolidated model order reduction techniques to solve parametric problems.
\end{itemize}


\section*{Acknowledgments}
We would like to acknowledge Giulia Campaniello who carried out the first numerical assessments in~\cite{giulia} under the supervision of the three authors.
This research is part of the activity of the METAMatLab at Politecnico di Milano.
\\
Nicola Ferro thanks Istituto Nazionale di Alta Matematica (INdAM) for the awarded grant.
Simona Perotto thanks the PRIN research grant n.20204LN5N5 \textit{Advanced Polyhedral Discretisations of Heterogeneous PDEs for Multiphysics Problems} and the INdAM-GNCS 2022 Project \textit{Metodi di riduzione computazionale per le scienze applicate: focus su sistemi complessi}. Matteo Gavazzoni acknowledges the Italian Ministry of University and Research for the support provided through the \textit{Department of Excellence LIS4.0 - Lightweight and Smart Structures for Industry 4.0} Project.

%

\bibliographystyle{elsarticle-num}
\bibliography{topoptbiblio.bib}

\begin{thebibliography}{10}
\expandafter\ifx\csname url\endcsname\relax
  \def\url#1{\texttt{#1}}\fi
\expandafter\ifx\csname urlprefix\endcsname\relax\def\urlprefix{URL }\fi
\expandafter\ifx\csname href\endcsname\relax
  \def\href#1#2{#2} \def\path#1{#1}\fi

\bibitem{Thompson2016}
M.~K. Thompson, G.~Moroni, T.~Vaneker, G.~Fadel, R.~I. Campbell, I.~Gibson,
  A.~Bernard, J.~Schulz, P.~Graf, B.~Ahuja, F.~Martina, Design for additive
  manufacturing: trends, opportunities, considerations, and constraints, CIRP
  Annals 65~(2) (2016) 737--760.

\bibitem{andreassen2014determine}
E.~Andreassen, C.~S. Andreasen, How to determine composite material properties
  using numerical homogenization, Comp. Mater. Sci. 83 (2014) 488--495.

\bibitem{allaire19}
G.~Allaire, P.~Geoffroy-Donders, O.~Pantz, Topology optimization of modulated
  and oriented periodic microstructures by the homogenization method, Comput.
  Math. Appl. 78~(7) (2019) 2197–2229.

\bibitem{sigmund1994}
O.~Sigmund, Materials with prescribed constitutive parameters: an inverse
  homogenization problem, Internat. J. Solids Structures 31~(17) (1994)
  2313--2329.

\bibitem{Rodrigues2002}
H.~Rodrigues, J.~Guedes, M.~Bendsoe, Hierarchical optimization of material and
  structure, Struct. Multidiscip. Optim. 24~(1) (2002) 1–10.

\bibitem{Sanders2021}
E.~D. Sanders, A.~Pereira, G.~H. Paulino, Optimal and continuous multilattice
  embedding, Sci. Adv. 7~(16) (2021) eabf4838.

\bibitem{ArabnejadKhanoki2012}
S.~Arabnejad~Khanoki, D.~Pasini, Multiscale design and multiobjective
  optimization of orthopedic hip implants with functionally graded cellular
  material, J. Biomech. Eng. 134~(3) (2012) 031004.

\bibitem{Radman2013}
A.~Radman, X.~Huang, Y.~M. Xie, Topology optimization of functionally graded
  cellular materials, J. Mater. Sci 48~(4) (2013) 1503–1510.

\bibitem{Panesar2018}
A.~Panesar, M.~Abdi, D.~Hickman, I.~Ashcroft, {Strategies for functionally
  graded lattice structures derived using topology optimisation for Additive
  Manufacturing}, Addit. Manuf. 19 (2018) 81--94.

\bibitem{Gao2019}
J.~Gao, Z.~Luo, H.~Li, L.~Gao, Topology optimization for multiscale design of
  porous composites with multi-domain microstructures, Comput. Methods Appl.
  Mech. Engrg. 344 (2019) 451--476.

\bibitem{coelho08}
P.~G. Coelho, P.~R. Fernandes, J.~M. Guedes, H.~C. Rodrigues, A hierarchical
  model for concurrent material and topology optimisation of three-dimensional
  structures, Struct. Multidiscip. Optim. 35~(2) (2008) 107–115.

\bibitem{Xia2014}
L.~Xia, P.~Breitkopf, {Concurrent topology optimization design of material and
  structure within FE2 nonlinear multiscale analysis framework}, Comput.
  Methods Appl. Mech. Eng. 278 (2014) 524--542.

\bibitem{Du2018}
Z.~Du, X.-Y. Zhou, R.~Picelli, H.~A. Kim, {Connecting microstructures for
  multiscale topology optimization with connectivity index constraints}, J.
  Mech. Des. 140~(11) (2018) 111417.

\bibitem{cramer2016}
A.~D. Cramer, V.~J. Challis, A.~P. Roberts, Microstructure interpolation for
  macroscopic design, Struct. Multidiscip. Optim. 53~(3) (2016) 489--500.

\bibitem{wang2017}
Y.~Wang, F.~Chen, M.~Y. Wang, Concurrent design with connectable graded
  microstructures, Comput. Methods Appl. Mech. Engrg. 317 (2017) 84--101.

\bibitem{zhou2008}
S.~Zhou, Q.~Li, Design of graded two-phase microstructures for tailored
  elasticity gradients, J. Mater. Sci. 43~(15) (2008) 5157–5167.

\bibitem{li2018}
H.~Li, Z.~Luo, L.~Gao, P.~Walker, Topology optimization for functionally graded
  cellular composites with metamaterials by level sets, Comput. Methods Appl.
  Mech. Engrg. 328 (2018) 340--364.

\bibitem{zhou2019}
X.-Y. Zhou, Z.~Du, H.~A. Kim, A level set shape metamorphosis with mechanical
  constraints for geometrically graded microstructures, Struct. Multidiscip.
  Optim. 60~(1) (2019) 1--16.

\bibitem{zobaer2020}
S.~M.~T. Zobaer, A.~Sutradhar, An energy-based method for interface
  connectivity of incompatible microstructures through parametric modeling,
  Comput. Methods Appl. Mech. Engrg. 370 (2020) 113278, 30.

\bibitem{liu2022}
X.~Liu, L.~Gao, M.~Xiao, Y.~Zhang, Kriging-assisted design of functionally
  graded cellular structures with smoothly-varying lattice unit cells, Comput.
  Methods Appl. Mech. Engrg. 390 (2022) 114466, 27.

\bibitem{schumacher2015}
C.~Schumacher, B.~Bickel, J.~Rys, S.~Marschner, C.~Daraio, M.~Gross,
  Microstructures to control elasticity in {3D} printing, {ACM} Trans. Graph.
  34~(4) (2015) 1--13.

\bibitem{garner2019}
E.~Garner, H.~M. Kolken, C.~C. Wang, A.~A. Zadpoor, J.~Wu, Compatibility in
  microstructural optimization for additive manufacturing, Addit. Manuf. 26
  (2019) 65--75.

\bibitem{liu2020}
P.~Liu, Z.~Kang, Y.~Luo, Two-scale concurrent topology optimization of lattice
  structures with connectable microstructures, Addit. Manuf. 36 (2020) 101427.

\bibitem{Borrvall2003}
T.~Borrvall, J.~Petersson, Topology optimization of fluids in {S}tokes flow,
  Internat. J. Numer. Methods Fluids 41~(1) (2003) 77--107.

\bibitem{Micheletti2019}
S.~Micheletti, S.~Perotto, L.~Soli, {Topology optimization driven by
  anisotropic mesh adaptation: Towards a free-form design}, Comput. \&
  Structures 214 (2019) 60--72.

\bibitem{ferro2020density}
N.~Ferro, S.~Micheletti, S.~Perotto, Density-based inverse homogenization with
  anisotropically adapted elements, in: A.~Corsini, S.~Perotto, G.~Rozza,
  H.~van Brummelen (Eds.), Numerical Methods for Flows, Vol. 132 of Lect. Notes
  Comput. Sci. Eng., Springer Cham, 2020, pp. 211--221.

\bibitem{ern04}
A.~Ern, J.-L. Guermond, Theory and Practice of Finite Elements,
  Springer-Verlag, New York, 2004.

\bibitem{Bendsoe2004}
M.~P. Bends{\o}e, O.~Sigmund, Topology Optimization, Springer, Heidelberg,
  Berlin, 2004.

\bibitem{brezzi2013}
D.~Boffi, F.~Brezzi, M.~Fortin, Mixed finite element methods and applications,
  Vol.~44 of Springer Series in Computational Mathematics, Springer,
  Heidelberg, 2013.

\bibitem{Ferro2020a}
N.~Ferro, S.~Micheletti, S.~Perotto, {Compliance–stress constrained mass
  minimization for topology optimization on anisotropic meshes}, SN Appl. Sci.
  2~(7) (2020) 1--11.

\bibitem{Ferro2020b}
N.~Ferro, S.~Micheletti, S.~Perotto, {An optimization algorithm for automatic
  structural design}, Comput. Methods Appl. Mech. Eng. 372 (2020) 113335.

\bibitem{Ferro2021}
N.~Ferro, S.~Perotto, D.~Bianchi, R.~Ferrante, M.~Mannisi, Design of cellular
  materials for multiscale topology optimization: application to
  patient-specific orthopedic devices, Struct. Multidiscip. Optim. 65~(3)
  (2022) 79.

\bibitem{Cristofaro2021}
D.~di~Cristofaro, C.~Galimberti, D.~Bianchi, R.~Ferrante, N.~Ferro, M.~Mannisi,
  S.~Perotto, {Adaptive topology optimization for innovative 3D printed
  metamaterials}, in: 14th World Congress on Computational Mechanics (WCCM) -
  Modeling and Analysis of Real World and Industry Applications, Vol. 1200,
  2020.

\bibitem{ferro19}
N.~Ferro, S.~Micheletti, S.~Perotto, {POD}-assisted strategies for structural
  topology optimization, Comput. Math. Appl. 77~(10) (2019) 2804--2820.

\bibitem{Frey2008}
P.~J. Frey, P.-L. George, Mesh Generation. Application to Finite Elements, 2nd
  Edition, ISTE, London; John Wiley \& Sons, Inc., Hoboken, NJ, 2008.

\bibitem{enumath09}
S.~Micheletti, S.~Perotto, Anisotropic adaptation via a {Z}ienkiewicz-{Z}hu
  error estimator for 2{D} elliptic problems, in: G.~Kreiss, P.~L{\"o}tstedt,
  A.~M{\aa}lqvist, M.~Neytcheva (Eds.), Numerical Mathematics and Advanced
  Applications, Springer-Verlag Berlin Heidelberg, 2010, pp. 645--653.

\bibitem{ZZ1987}
O.~C. Zienkiewicz, J.~Z. Zhu, A simple error estimator and adaptive procedure
  for practical engineerng analysis, Int. J. Numer. Methods Eng. 24~(2) (1987)
  337--357.

\bibitem{Zienkiewicz1992}
O.~C. Zienkiewicz, J.~Z. Zhu, The superconvergent patch recovery and a
  posteriori error estimates. {II}: {E}rror estimates and adaptivity, Int. J.
  Numer. Meth. Engng 33 (1992) 1365--1382.

\bibitem{Dompierre2002}
J.~Dompierre, M.-G. Vallet, Y.~Bourgault, M.~Fortin, W.~G. Habashi, Anisotropic
  mesh adaptation: towards user-independent, mesh-independent and
  solver-independent {CFD}. {III}. {U}nstructured meshes, Internat. J. Numer.
  Methods Fluids 39~(8) (2002) 675--702.

\bibitem{perotto2015}
S.~Perotto, L.~Formaggia (Eds.), New Challenges in Grid Generation and
  Adaptivity for Scientific Computing, Springer International Publishing, 2015.

\bibitem{perotto2022}
R.~Sevilla, S.~Perotto, K.~Morgan (Eds.), Mesh Generation and Adaptation.
  Cutting-Edge Techniques., Springer Cham, 2022.

\bibitem{filtrati}
B.~S. Lazarov, O.~Sigmund, Filters in topology optimization based on
  {H}elmholtz-type differential equations, Int. J. Numer. Meth. Engng 86~(6)
  (2011) 765--781.

\bibitem{SP98}
O.~Sigmund, J.~Petersson, Numerical instabilities in topology optimization: a
  survey on procedures dealing with checkerboards, mesh-dependencies and local
  minima, Struct. Optim. 16~(1) (1998) 68--75.

\bibitem{Lions1971}
J.-L. Lions, Optimal Control of Systems Governed by Partial Differential
  Equations, Springer-Verlag, New York-Berlin, 1971.

\bibitem{margossian2019}
C.~C. Margossian, A review of automatic differentiation and its efficient
  implementation, {WIREs} Data Mining and Knowl. Discov. 9~(4) (Mar. 2019).

\bibitem{Bolla06}
S.~Micheletti, S.~Perotto, Reliability and efficiency of an anisotropic
  {Z}ienkiewicz-{Z}hu error estimator, Comput. Methods Appl. Mech. Engrg.
  195~(9-12) (2006) 799--835.

\bibitem{Micheletti2010}
S.~Micheletti, S.~Perotto, P.~E. Farrell, A recovery-based error estimator for
  anisotropic mesh adaptation in {CFD}, Bol. Soc. Esp. Mat. Apl. SeMA 50 (2010)
  115--137.

\bibitem{Gavazzoni2022}
M.~Gavazzoni, N.~Ferro, S.~Perotto, S.~Foletti, Multi-physics inverse
  homogenization for the design of innovative cellular materials: Application
  to thermo-elastic problems, Math. Comput. Appl. 27~(1) (2022) 15.

\bibitem{Waechter2006}
A.~W\"achter, L.~T. Biegler, On the implementation of an interior-point filter
  line-search algorithm for large-scale nonlinear programming, Math. Program.
  106~(1, Ser. A) (2006) 25--57.

\bibitem{FreeFem}
F.~Hecht, New development in {F}ree{F}em++, J. Numer. Math. 20~(3-4) (2012)
  251--265.

\bibitem{li2016}
Q.~Li, W.~Chen, S.~Liu, L.~Tong, Structural topology optimization considering
  connectivity constraint, Struct. Multidiscip. Optim. 54~(4) (2016) 971--984.

\bibitem{giulia}
G.~Campaniello, Topology optimization for the design of manufacturable cellular
  materials, Master's thesis, Politecnico di Milano (2021).

\end{thebibliography}


\end{document}